\documentclass[a4paper,11pt]{article}
\usepackage[left=20mm,right=20mm,top=20mm,bottom=20mm]{geometry}
\usepackage{amssymb, amsmath, amsthm, mathtools}
\usepackage{bm}
\usepackage{mathrsfs}
\usepackage{hyperref,bookmark}
\usepackage{graphicx}
\usepackage{epsfig,color}
\usepackage{mathrsfs}
\usepackage{verbatim}
\usepackage{url}
\usepackage{subcaption}
\usepackage{float}
\usepackage[greek,english]{babel}
\usepackage{dcolumn}
\usepackage{accents}
\usepackage{cases}
\newcommand{\tr}{\operatorname{tr}}
\newcommand{\dd}{\operatorname{d}\!}

\newcommand{\n}{\bm{n}}
\newcommand{\m}{\bm{m}}
\newcommand{\e}{\bm{e}}

\newcommand{\body}{\mathscr{B}}
\newcommand{\free}{\mathscr{F}}

\newcommand{\vv}{\bm{v}}

\newcommand{\diam}{\mathrm{diam}\left(\body_0\right)}
\newcommand{\bF}{\bm{\mathrm{F}}}
\newcommand{\bL}{\bm{\mathrm{L}}}
\newcommand{\bG}{\bm{\mathrm{G}}}

\newcommand{\bP}{\bm{\mathrm{P}}}

\theoremstyle{definition}

\begin{document}

\title{\huge\textsf{\textbf{Shape instabilities driven by topological defects in nematic polymer networks}}}
\author{Silvia Paparini\footnote{Corresponding author.} $^\dagger$ \qquad
	Giulio G. Giusteri\footnote{Dipartimento di Matematica “Tullio Levi-Civita”, Università degli Studi di Padova, Padua, Italy, and Gruppo Nazionale per la Fisica Matematica, Istituto Nazionale di Alta Matematica “Francesco Severi”. Email: \texttt{silvia.paparini@unipd.it}, \texttt{giulio.giusteri@unipd.it}}\qquad
L. Angela Mihai\footnote{School of Mathematics, Cardiff University, Cardiff, UK, Email: \texttt{MihaiLA@cardiff.ac.uk}.}}
\date{}
\maketitle

\begin{abstract}\vskip 6pt
\hrule\vskip 12pt
Liquid crystalline networks (LCNs) are stimuli-responsive materials formed from polymeric chains cross-linked with rod-like mesogenic segments, which, in the nematic phase, align along a non-polar director. A key characteristic of these nematic systems is the existence of singularities in the director field, known as topological defects or disclinations, and classified by their topological charge. In this study, we address the open question of modeling theoretically the coupling between mesogens disclination and polymeric network by providing a mathematical framework describing the out-of-plane shape changes of initially flat LCN sheets containing a central topological defect. Adopting a variational approach, we define an energy associated with the deformations consisting of two contributions: an elastic energy term accounting for spatial director variations, and a strain-energy function describing the elastic response of the polymer network. The interplay between nematic elasticity, which seeks to minimize distortions in the director field, variations in the degree of order, with the consequent tendency of monomers in the polymer chains to distribute anisotropically in response to an external stimulus, and mechanical stiffness, which resists deformation, determines the resulting morphology. We analyze the transition to instability of the ground-state flat configuration and characterize the corresponding buckling modes. \\



\vskip 12pt
\hrule
\end{abstract}

\section{Introduction}

Nematic liquid crystal networks (LCNs) are anisotropic materials combining the properties of polymeric chains and nematic liquid crystals (LCs) \cite{McCracken:2020:etal,Mihai:2022,Terentjev:2025,Warner:2007:WT,White:2018,White:2015:WB,Xiao:etal:2024}. {\color{black}Similarly to classical LCs, in LCNs, the mesogenic segments in the nematic phase possess orientational order but lack positional ordering of their centers of mass. Moreover, different from liquid crystal elastomers (LCEs), in LCNs, cross-linking is tight and mobility of the nematic director is limited with respect to the polymer network, hence the coupling between nematic order and polymer is strong enough to constrain the director field to follow material deformation \cite{Warner:2020}.} A significant factor in the mechanical behavior of LCNs is the presence of \emph{topological defects} or \emph{disclinations} \cite{Selinger:2024}. These are singularities in the nematic orientation, which emerge when LC molecules encounter sudden geometrical changes, such as sharp edges or corners, and can be classified by their topological \emph{charge} \cite{McConney:2013:etal}.

Many biological systems of cells and cytoskeletal elements can also form a nematic phase where elongated constituents align parallel to each other, inducing partial orientational order similar to that observed in nematic LCs \cite{blanch:turbulent,keber:topology,kumar:tunable,narayan:long,saw:topological}. For these systems, topological defects in the nematic order can act as organizing centers enabling organisms to grow protrusions or deplete material to relieve stress \cite{copenhagen:topological,kawaguchi:topological,keber:topology,Maroudas-Sacks:2021:etal,saw:topological}. In Hydra's ectoderm, for example, topological defects align with morphological features: a defect of charge $+1$ is localized at the tentacle's tip, and two $-1/2$ defects reside at its base \cite{Maroudas-Sacks:2021:etal}.

{\color{black} While LC disclinations and textures are ubiquitous in natural and synthetic soft matter, their coupling with the polymeric network in LCNs is a difficult task to model analytically, especially for topological charges different from $+1$ \cite{Fried:2001:FT,Fried:2002:FT,Mihai:2020b:MG}. Defects of charge $+1$ have been primarily studied, for example, in \cite{fried2002:normal,modes2011:gaussian,modes2010:disclination}. Emergent shapes with Gaussian curvature localized at point defects are discussed in \cite{modes2011:gaussian,Warner:2007:WT}, while numerical approaches such as shell theory simulations \cite{duffy2020:defective} and finite element methods with regularization \cite{bouck2024:reduced} have been employed to capture richer physical phenomena in LCNs, including origami-like structures and deformations due to defects with varying topological charges. 

In this work, we develop a mathematical framework for the study of LCN defects by building on results from the LC theory \cite{Long:2021:etal,paparini:shape,paparini:spiralling,paparini:stability,Tang:2017:TS}. The key characteristic of our approach, which also distinguishes our model from, for example \cite{modes2011:gaussian,duffy2020:defective,bouck2024:reduced}, in the treatment of defects with different topological charges, is that we consider a regime in which the nematic phase around the defect is well established, and shape deformations arise to relieve the mechanical stress originating from the imprinted, distorted director field with topological disclinations, as well as from variations in the degree of order and mechanical stiffness. Accordingly, an elastic energy that penalizes spatial distortions in the director field, measured by its gradient, is employed, along with a strain-energy function that describes the elastic response of the nematic polymer network to deformations, depending on the degrees of orientation and the director fields before and after the application of an external stimulus. This framework is inspired by biological shape formation and morphogenesis, as well as by the shape morphing of LCN sheets studied in \cite{McConney:2013:etal}. Within this formalism, we adopt a variational approach that enables an almost entirely analytical solution to the out-of-plane buckling problem in nematic polymer networks.}

In the following sections, we first define topological defects in planar LCN sheets (Section~\ref{sec:defects}). We then describe mathematically the formation of topological defects caused by physical changes when external stimuli are applied (Section~\ref{sec:model}). To demonstrate our theoretical framework, we present computed examples showing out-of-plane shape deformations in initially flat LCN samples similar to those seen in physical experiments (Section~\ref{sec:examples}). {\color{black} In this context, the thickness of the flat LCN sheet is assumed to be sufficiently small to ensure that the imprinted director field has no significant component or variation in the thickness direction, and can therefore be extended uniformly across the cross-section. Our energy will scale linearly with the thickness and account for stretching. The resulting model will be therefore a membrane model \cite{cesana2015:effective,ozenda2020:blend,conti2018:adaptive}. Another popular scaling of the free energy with the thickness is cubic, leading to a plate model governed by bending effects. Examples include models derived via formal asymptotics \cite{ozenda2020:blend} and a von K\'{a}rm\'{a}n plate model obtained in \cite{Mihai:2020b:MG} using asymptotic analysis.} In the concluding section, we highlight key challenges in deriving our results and provide an outlook on further investigations. More substantial detailed calculations are deferred to appendices.

\section{Topological defects in the plane}\label{sec:defects}
 
Nematic order in LCNs is described by a non-polar unit vector $\n\in\mathbb{S}^2$, called the \emph{director}, and a \emph{scalar order parameter} $s\in [0, 1]$. The former describes the local average orientation of the LC molecules at the macroscopic scale, and has a bearing on the spatial organization of polymer strands. {\color{black} The constitutive molecules of nematic liquid crystals do not possess a permanent dipole moment and exhibit head-tail symmetry, meaning that the director field $\n$ is physically equivalent to $-\n$.}
The latter represents the average orientation of the nematogenic constituents at the microscopic scale, with $s=1$ when LC molecules are perfectly aligned with each other, and $s=0$ when the material becomes isotropic. {\color{black} According to Ericksen’s theory \cite{ericksen:liquid}, whose formalism will be employed in the following section, defects correspond to localized isotropic regions where the scalar order parameter $s$ vanishes, reflecting the absence of a preferred molecular orientation. Spatial variations in $s$ can serve to relax the distortion energy associated with defects.} 

\begin{figure}[htb] 
	\centering
	\includegraphics[width=0.45\linewidth]{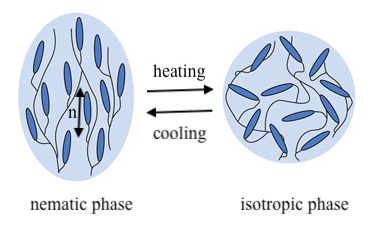}
	\caption{Thermally induced phase transition in LCN material (reprinted from \cite{gruzdenko2022:liquid}).}
	\label{fig:small_element}
\end{figure}

\begin{figure}[htbp] 
	\centering
	\includegraphics[width=0.6\linewidth]{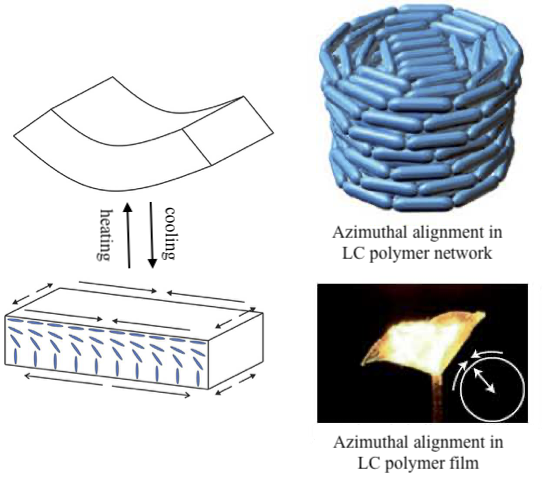}
	\caption{Thermally induced macroscopic shape changes in a LCN sheet containing a disclination (reprinted from \cite{fernandez2013:liquid,gruzdenko2022:liquid}).}
	\label{fig:shape_def_examples}
\end{figure}

For illustration, in a thermotropic mesogenic substance, the nematic phase is induced by changing its temperature, as shown in Figure~\ref{fig:small_element}. In the nematic phase, at a temperature lower than $\mathrm{T}_{\mathrm{c}}$ (nematic-isotropic transition temperature), mesogens and cross-linking sites are uniaxially oriented along $\n$, and the polymeric network is extended in the same direction. When heated over $\mathrm{T}_{\mathrm{c}}$, the LC molecules lose their orientational order, and the polymeric network contracts in the direction which was previously along $\n$. These deformations are local and reversible. Local deformations then drive macroscopic shape changes. For example, in Figure~\ref{fig:shape_def_examples}, the distribution of the director field is marked by blue ellipsoids. On the left, the bending of an LCN film with splay alignment is illustrated. This bending occurs due to thermally induced contraction of the top side and elongation of the bottom side along the same direction (the elongation in the orthogonal direction does not contribute to the bending). In the right panel, a macroscopic deformation guided by an azimuthal LC alignment is shown. In this case, the flat LCN film evolves into a nearly conical shape.

\begin{figure}[htbp]
	\centering
	\begin{subfigure}[c]{0.25\linewidth}
		\centering
		\includegraphics[width=0.75\linewidth]{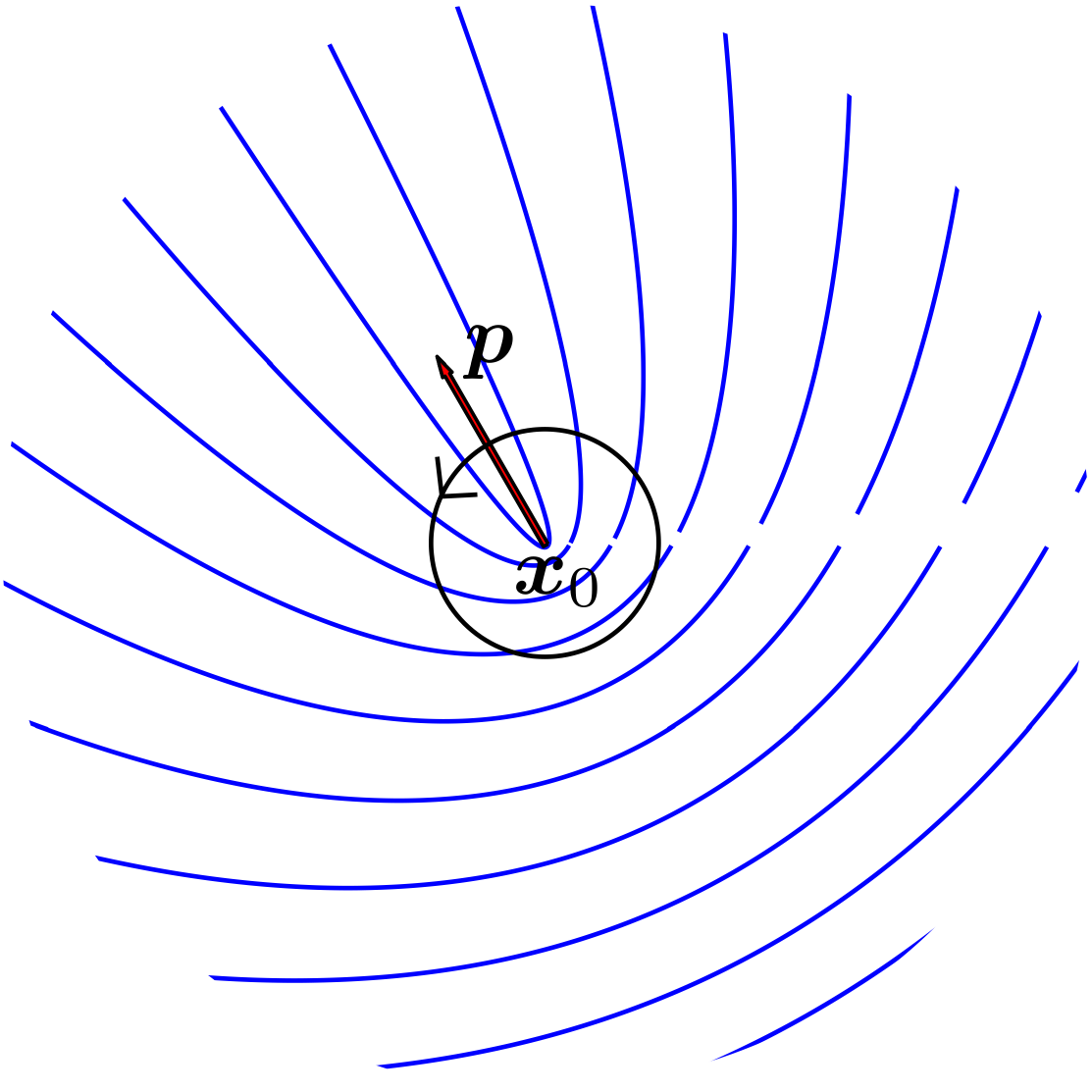}
		\caption{$q=+\dfrac{1}{2}$, $\vartheta_0=\dfrac{\pi}{3}$\\ One periodicity region.} 
		\label{fig:d_1o2}
	\end{subfigure}
	\begin{subfigure}[c]{0.25\linewidth}
		\centering
		\includegraphics[width=0.75\linewidth]{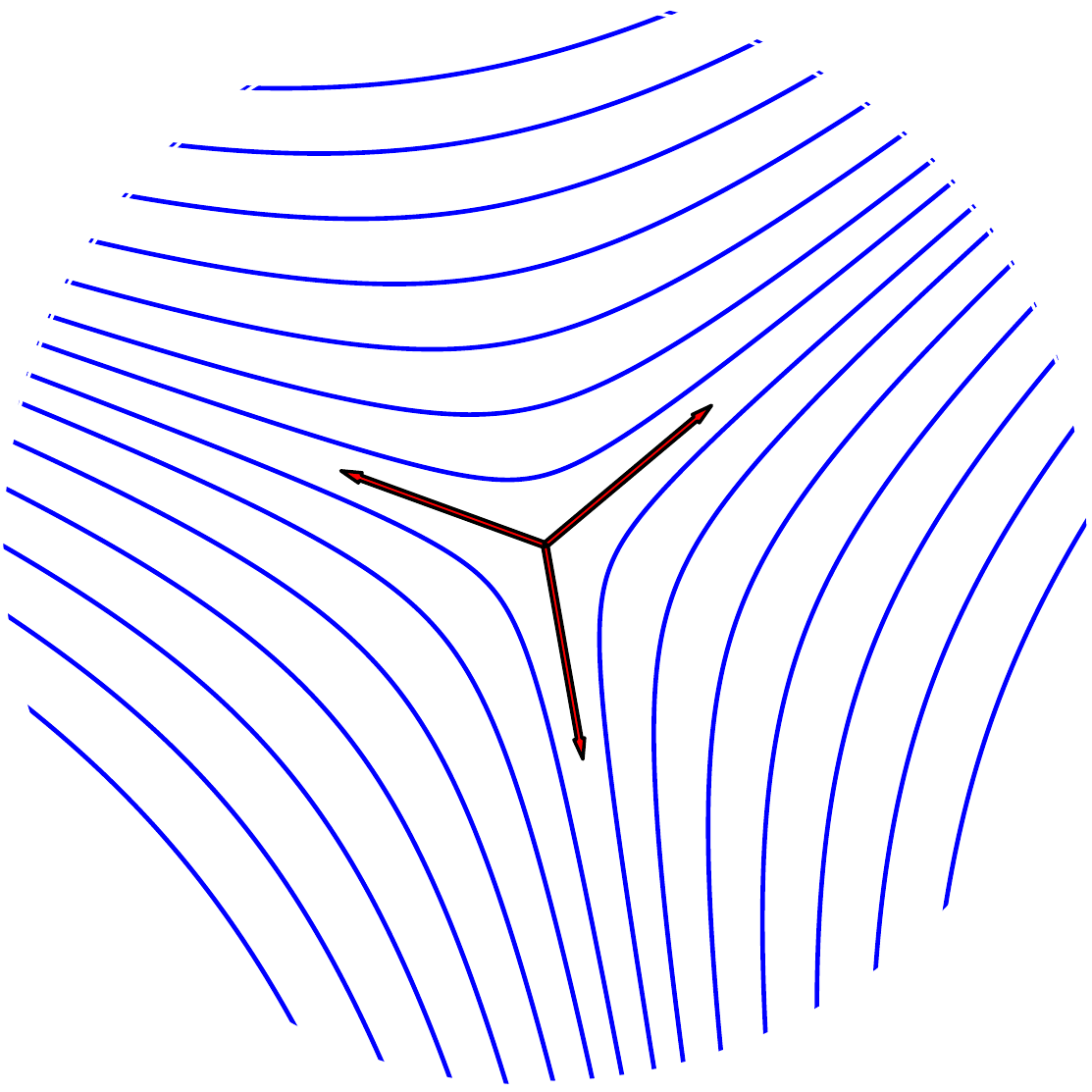}
		\caption{$q=-\dfrac{1}{2}$, $\vartheta_0=\dfrac{\pi}{3}$\\ Three periodicity regions} 
		\label{fig:d_minus_1o2}
	\end{subfigure}\\
		\begin{subfigure}[c]{0.25\linewidth}
		\centering
		\includegraphics[width=0.75\linewidth]{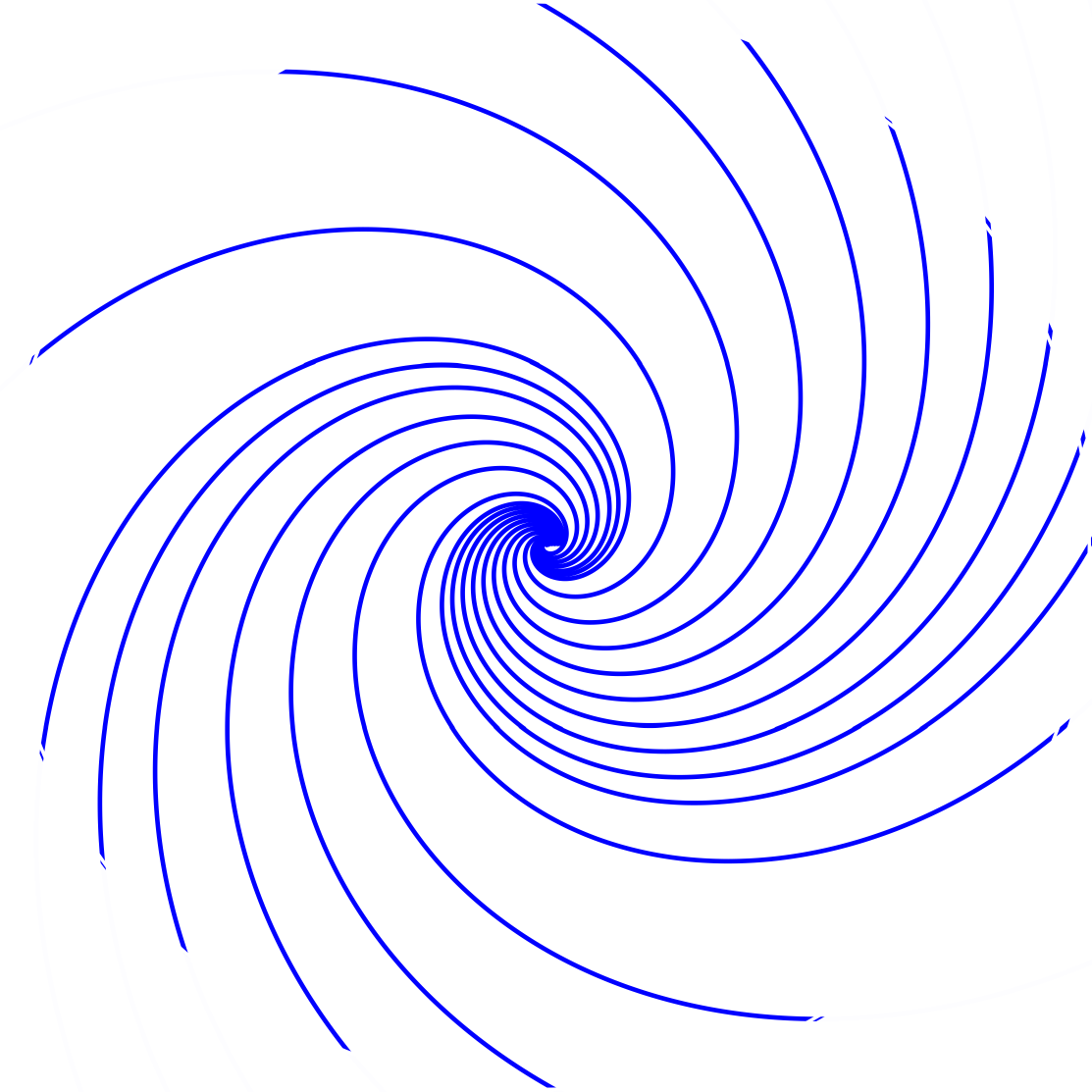}
		\caption{$q=+1$, $\vartheta_0=\dfrac{\pi}{3}$\\ Director points radially.}
		\label{fig:d_1}
	\end{subfigure}
	\begin{subfigure}[c]{0.25\linewidth}
		\centering
		\includegraphics[width=0.75\linewidth]{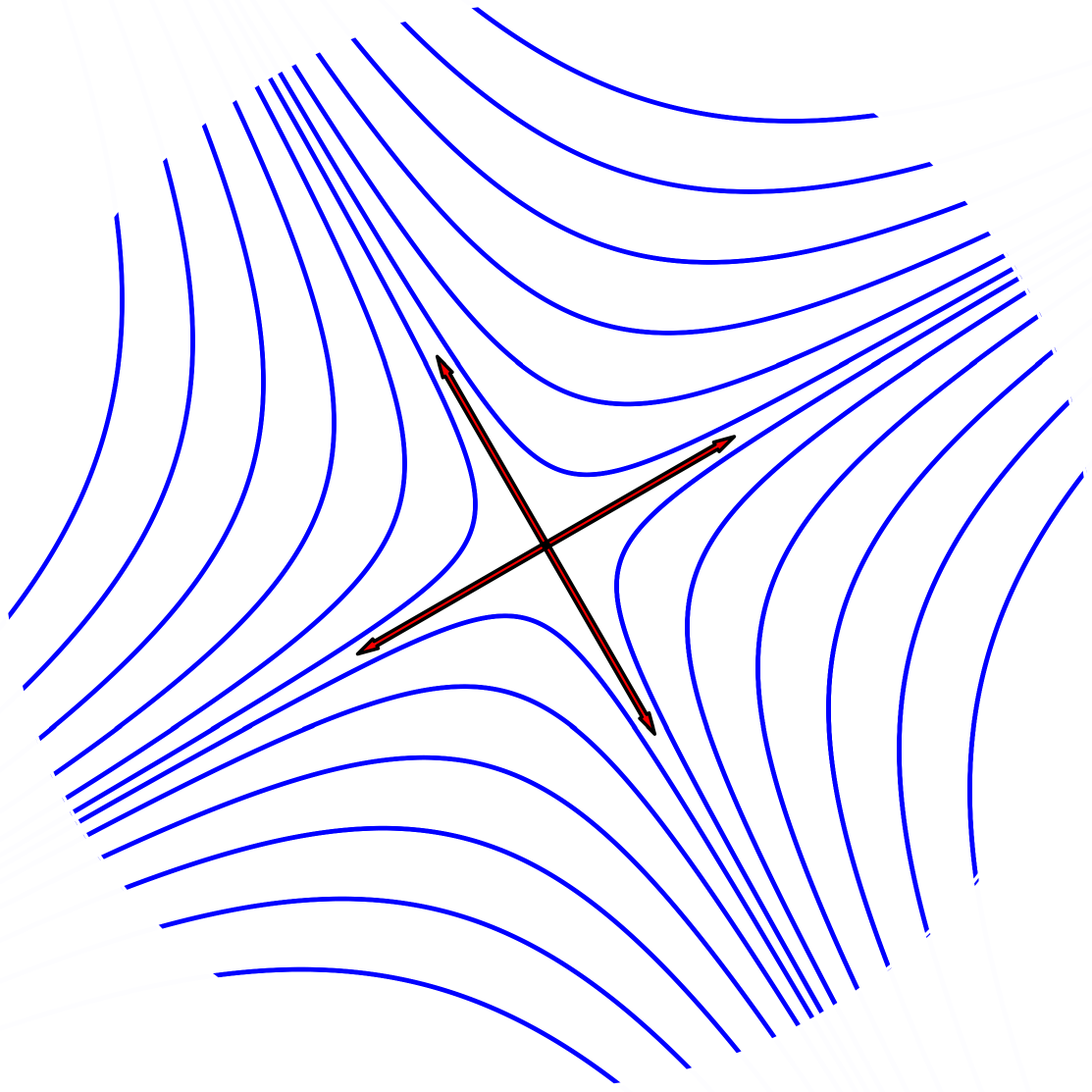}
		\caption{$q=-1$, $\vartheta_0=\dfrac{\pi}{3}$\\ Four periodicity regions.}
		\label{fig:d_minus_1}
	\end{subfigure}\\
	\begin{subfigure}[c]{0.25\linewidth}
		\centering
		\includegraphics[width=0.75\linewidth]{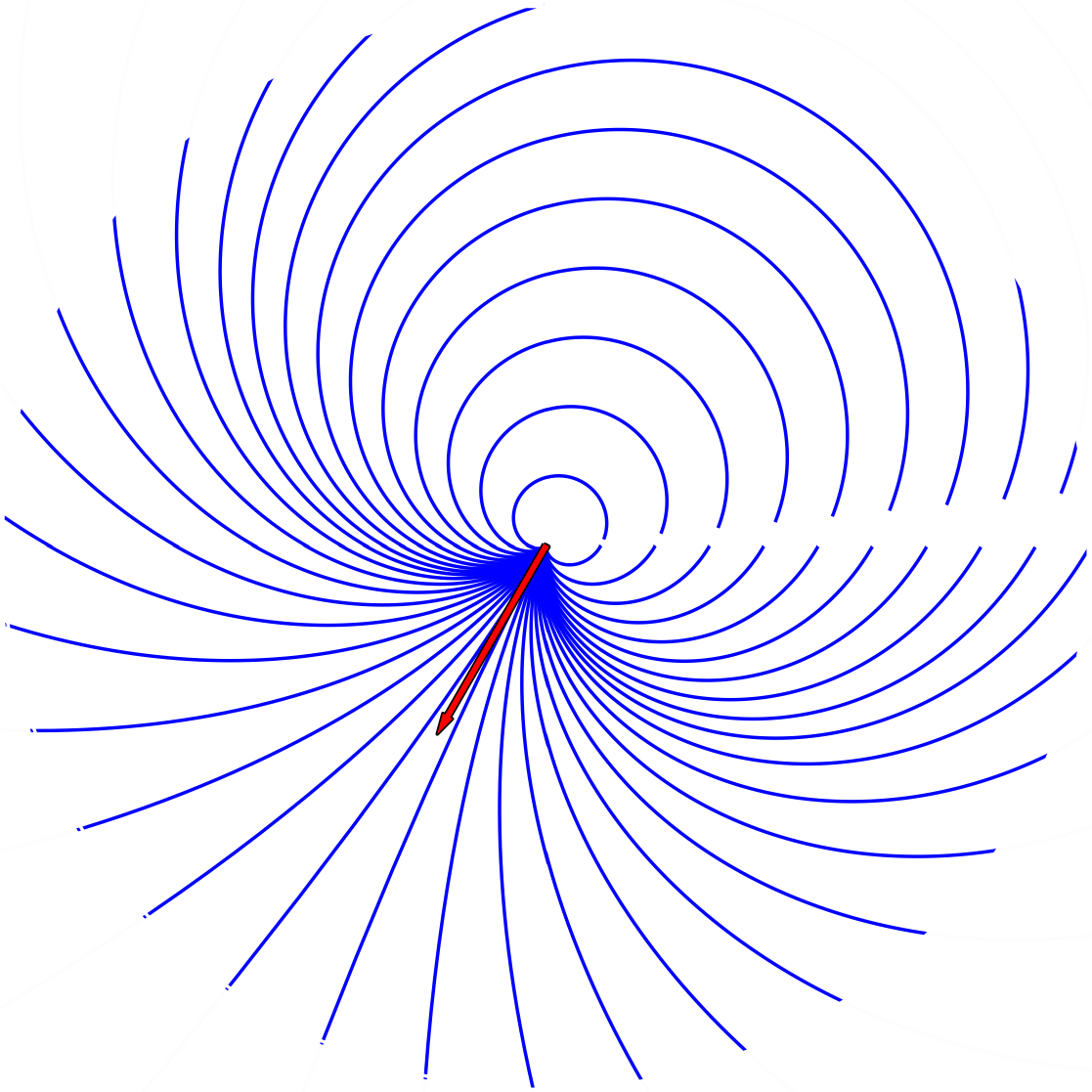}
		\caption{$q=+\dfrac{3}{2}$, $\vartheta_0=\dfrac{\pi}{3}$\\ One periodicity region.}
		\label{fig:d_3o2}
	\end{subfigure}
	\begin{subfigure}[c]{0.25\linewidth}
		\centering
		\includegraphics[width=0.75\linewidth]{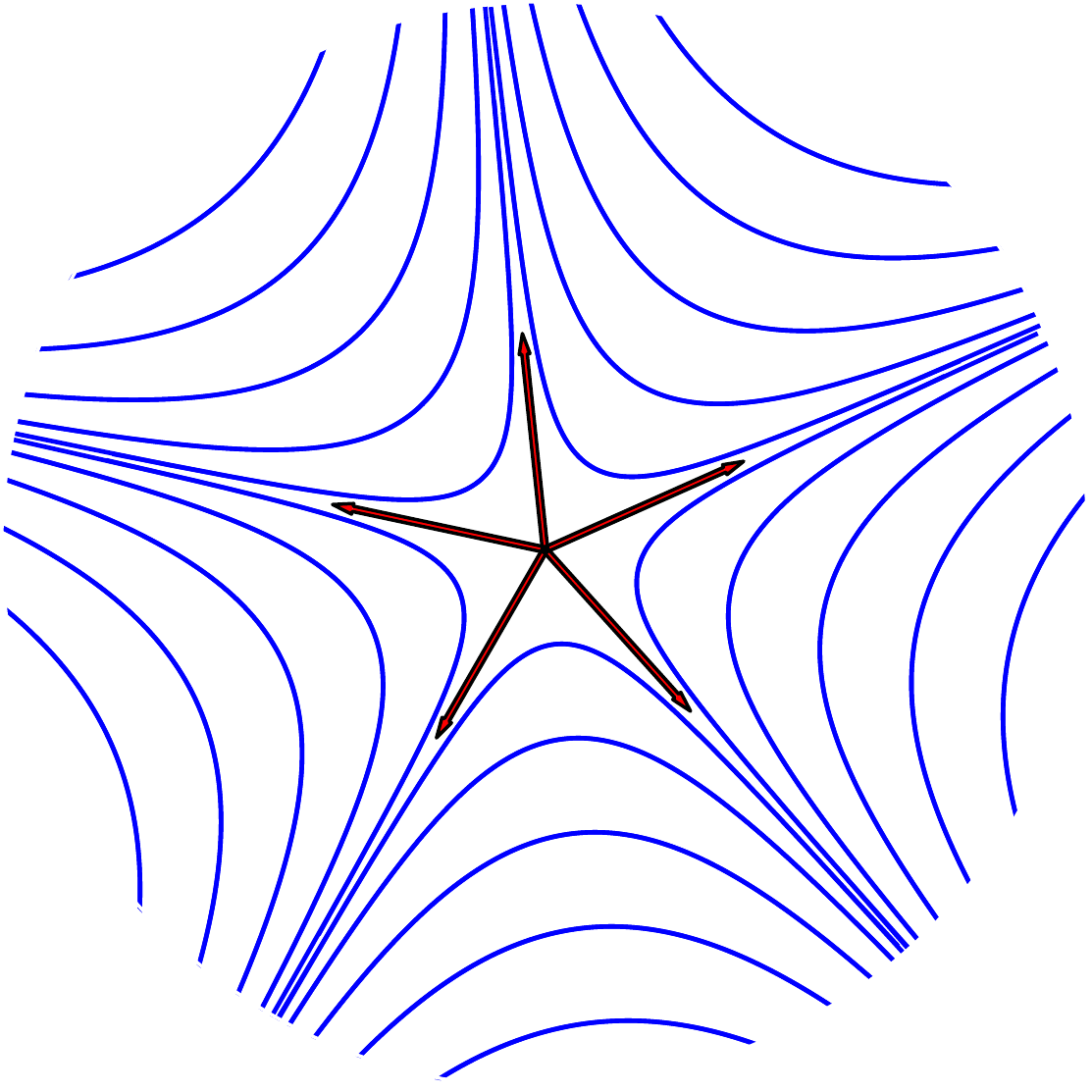}
		\caption{$q=-\dfrac{3}{2}$, $\vartheta_0=\dfrac{\pi}{3}$\\ Five periodicity regions.}
		\label{fig:d_minus_3o2}
	\end{subfigure}\\
	\begin{subfigure}[c]{0.25\linewidth}
		\centering
		\includegraphics[width=0.75\linewidth]{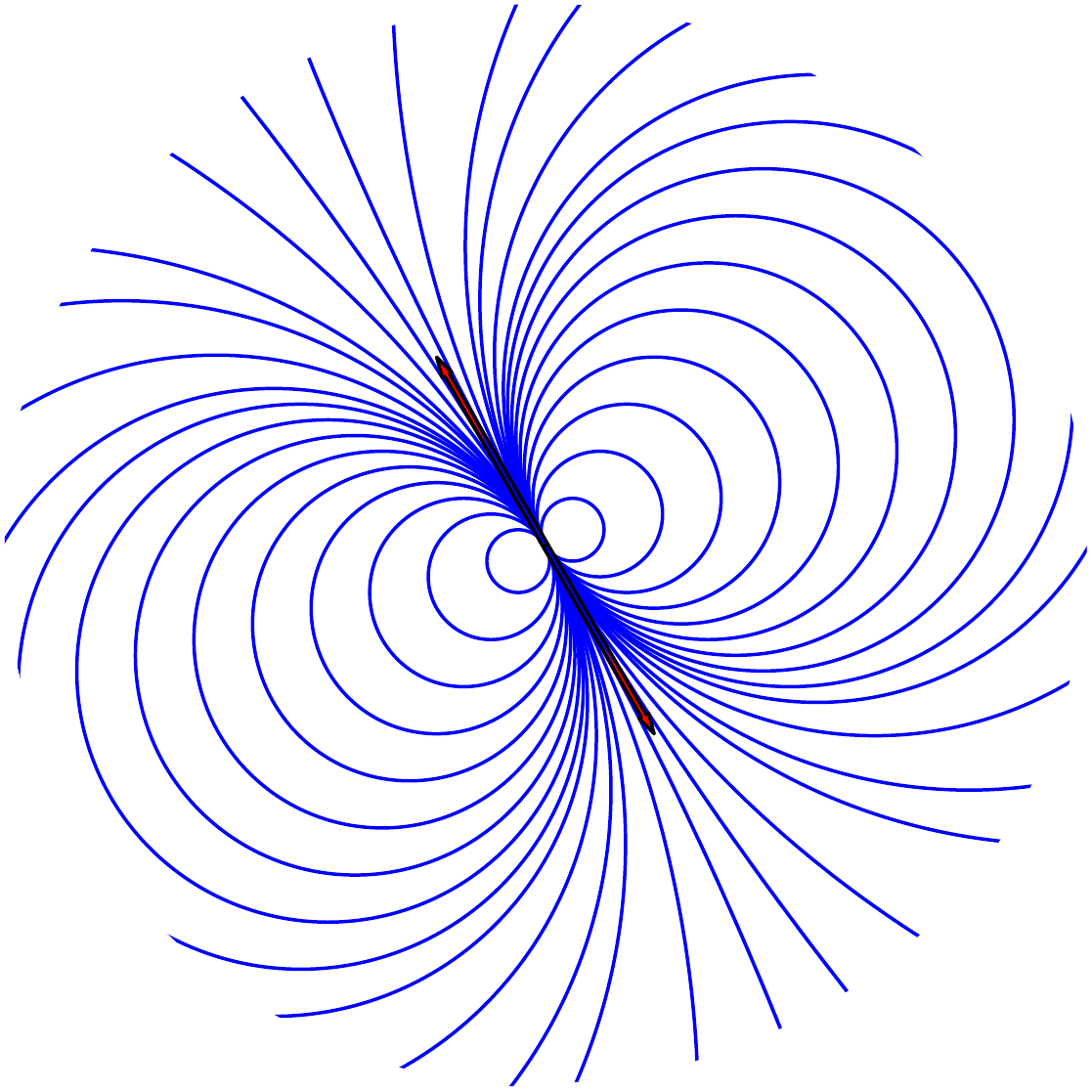}
		\caption{$q=+2$, $\vartheta_0=\dfrac{\pi}{3}$\\ Two periodicity regions.}
		\label{fig:d_2}
	\end{subfigure}
	\begin{subfigure}[c]{0.25\linewidth}
		\centering
		\includegraphics[width=0.75\linewidth]{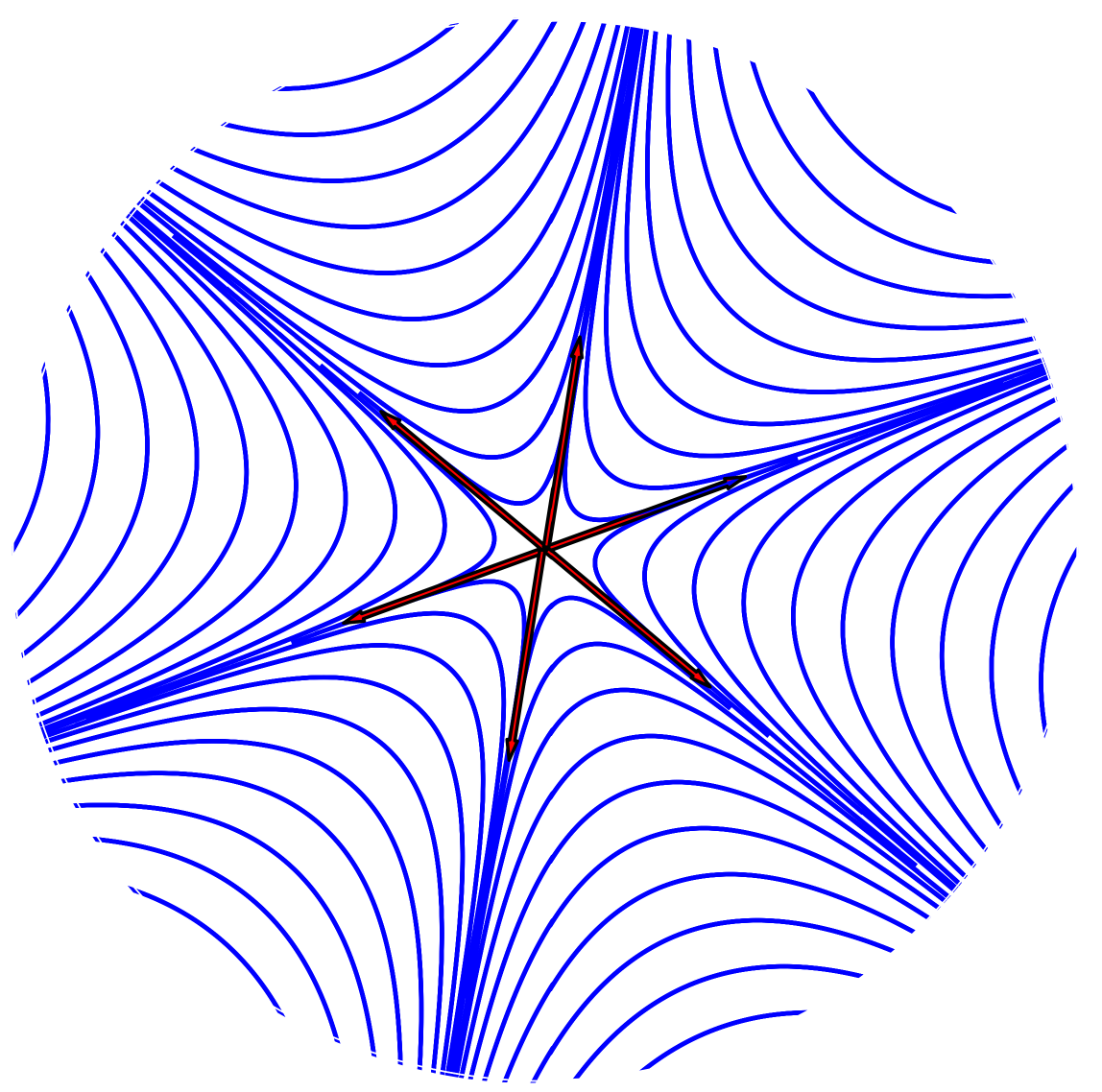}
		\caption{$q=-2$, $\vartheta_0=\dfrac{\pi}{3}$\\ Six periodicity regions.}
		\label{fig:d_minus_2}
	\end{subfigure}\\
    \begin{subfigure}[c]{0.25\linewidth}
		\centering
		\includegraphics[width=0.75\linewidth]{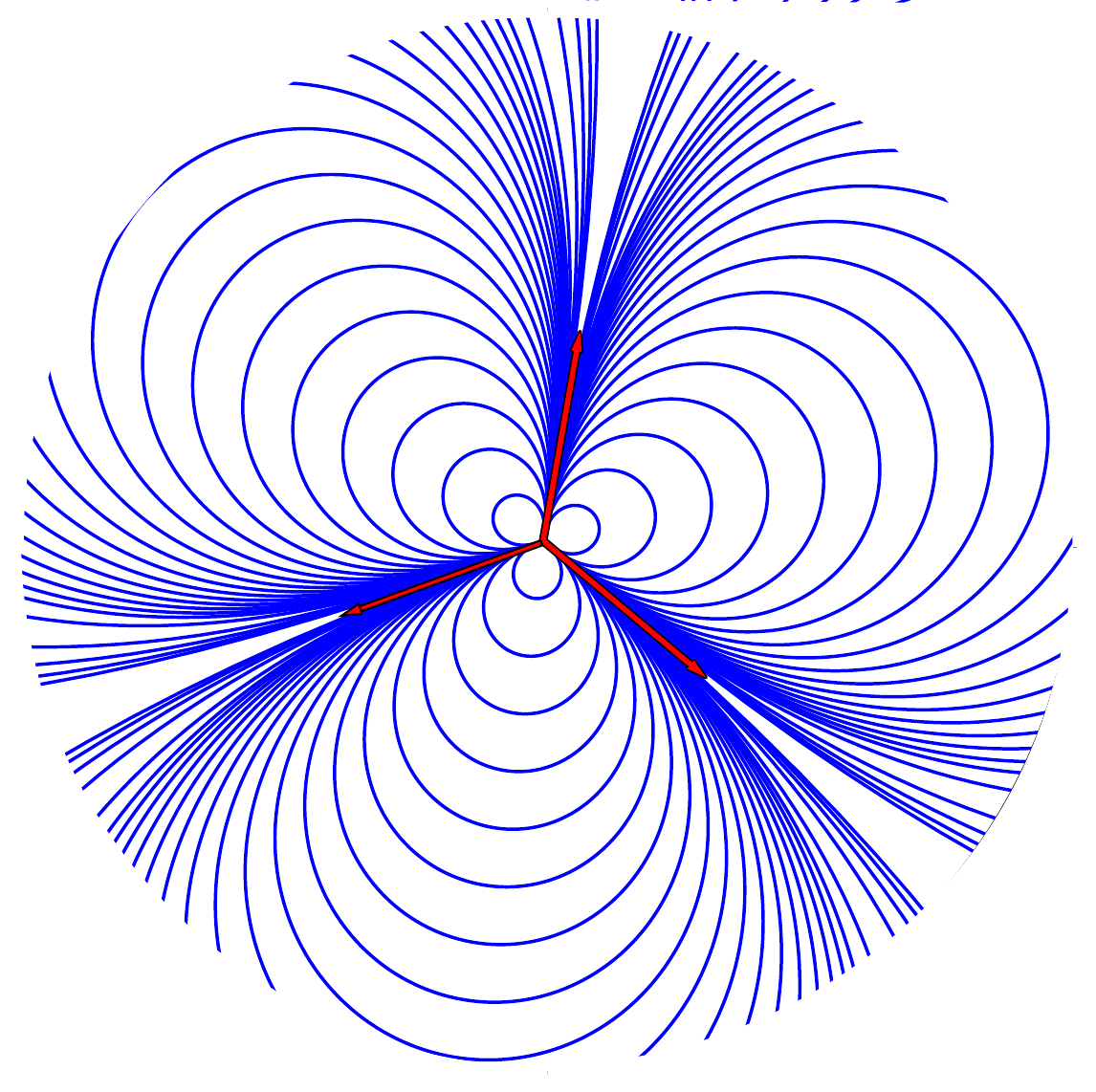}
		\caption{$q=+\dfrac{5}{2}$, $\vartheta_0=\dfrac{\pi}{3}$\\ Three periodicity regions.}
		\label{fig:d_5o2}
	\end{subfigure}
	\begin{subfigure}[c]{0.25\linewidth}
		\centering
		\includegraphics[width=0.75\linewidth]{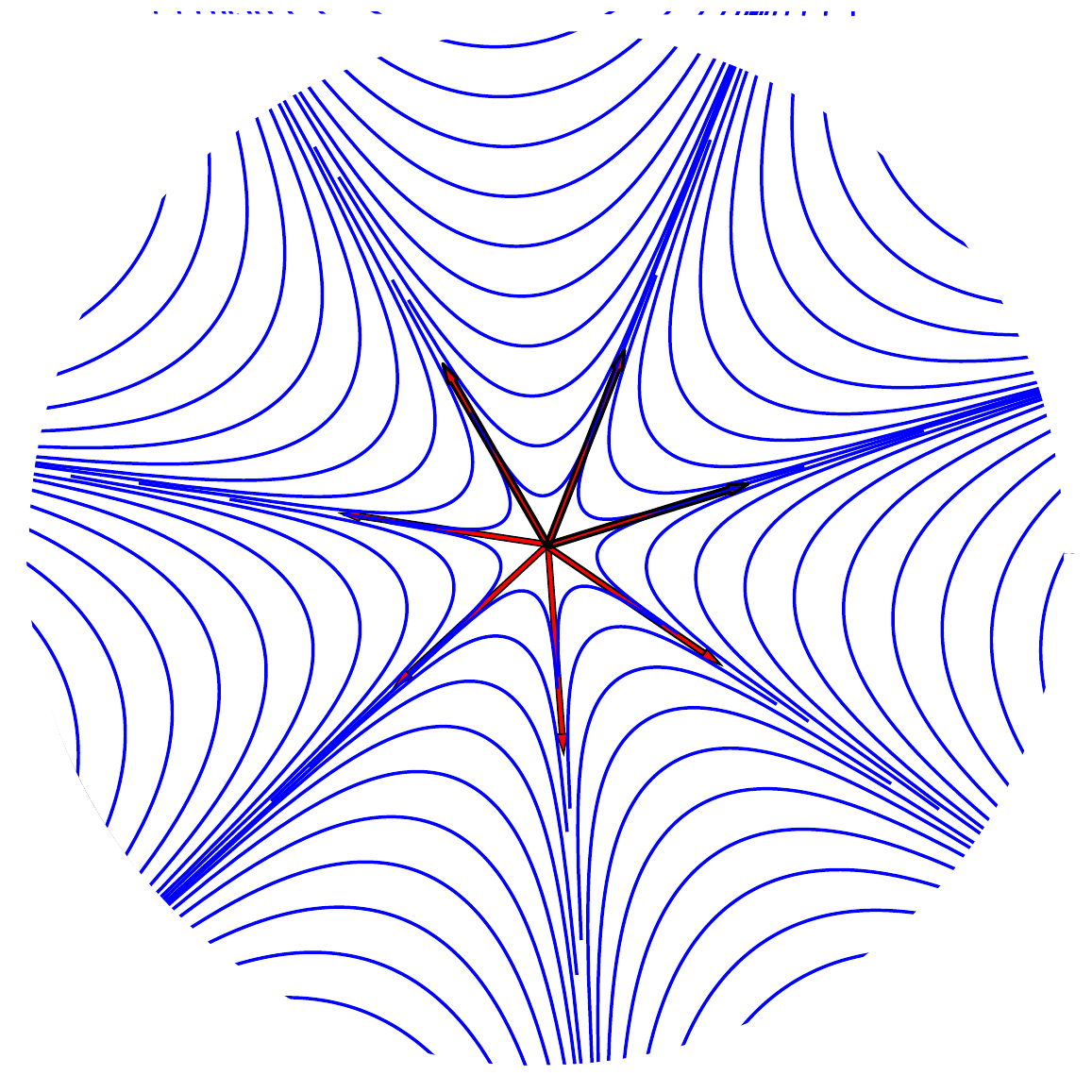}
		\caption{$q=-\dfrac{5}{2}$, $\vartheta_0=\dfrac{\pi}{3}$\\ Seven periodicity regions.}
		\label{fig:d_minus_5o2}
	\end{subfigure}
\caption{Field lines of $\m_0$ defined by \eqref{eq:n_0_q}, for a defect with topological charge $q$ at $\bm x_0$. Red arrows indicate the orientations $\bm p$ described by \eqref{eq:orientation_defect}, which identify the region of periodicity for each defect.}
	\label{fig:integral_lines_defects}
\end{figure}

Assuming a planar geometry, let a unit vector field $\m_0$ lie in the plane, and $\e_z$ be the unit outer normal vector to the plane. We define the \emph{topological charge} $q$ of $\m_0$, with a point defect at $\bm x_0$, as the \emph{winding number} of $\m_0$ on the tangent plane of the defect \cite{kralj:curvature}. Consider any simple circuit $\mathcal{C}$ within the plane, around $\bm x_0$, that can be continuously contracted toward $\bm x_0$. Conventionally, $\mathcal{C}$ has an anti-clockwise orientation around $\e_z$. Let $\e_0$ be a unit vector in the plane defined at a point along $\mathcal{C}$. The angle by which $\m_0$ rotates about $\e_0$ along the complete circuit $\mathcal{C}$ equals $2\pi q$, where $q$ is the topological charge of the defect. This is positive or negative, depending on whether the complete turn of $\m_0$ is consistent or not with the orientation of $\mathcal{C}$.

The director field with topological charge $q$ on a plane with a defect at the origin is represented with respect to a standard cylindrical coordinate system $(r,\vartheta,z)$ as
\begin{equation}
\label{eq:n_0_q}
\m_0=(\cos w) \e_r+(\sin w) \e_\vartheta, \qquad w=w(\vartheta)=(q-1)\vartheta+\vartheta_0.
\end{equation}
Here, $\e_r$ and $\e_\vartheta$ are the radial and azimuthal directions, respectively, $w=w(\vartheta)$ represents the angle formed by $\m_0$ with the radial direction $\e_r$, and $\vartheta_0$ is the phase of the defect, i.e., the arbitrary overall rotation of the director about the $z$-axis. 

To characterize the orientation of a defect {\color{black} when $q\neq1$}, we find when the director $\m_0$ points radially outward from (or inward toward) the defect, i.e., when $\m_0=\e_r$ \cite{Tang:2017:TS}. This happens whenever $w\equiv 0$, and the following vectors are identified
\begin{equation}
\label{eq:orientation_defect} 
\bm p=\cos\left(\frac{\vartheta_0}{1-q}-1\right)\e_r+\sin\left(\frac{\vartheta_0}{1-q}-1\right)\e_\vartheta \ \hbox{mod}\left(\frac{\vartheta_0}{|1-q|}\right)
\end{equation}
Accordingly, for a defect of topological charge {\color{black} $q\neq 1$}, the orientation vectors $\bm p$ identify $2|1-q|$ regions in which $\m_0$ and $-\m_0$ repeat themselves, as shown in Figure~\ref{fig:integral_lines_defects}. {\color{black} When $q=1$, $\m_0$ points always radially}.

We see from \eqref{eq:n_0_q} that the field lines of $\m_0$ are the solutions $(r(\tau),\vartheta(\tau))$ to the differential system
\begin{equation}
\label{eq:diff_system}
\frac{\dd \vartheta}{\dd\tau}=1,\qquad
\hat{r}'(\vartheta(\tau)))=\frac{r(\vartheta(\tau))}{\tan w(\vartheta(\tau))},
\end{equation}
subject to 
\begin{equation}
\vartheta(0)=\bar\vartheta  \ \hbox{mod}\left(\frac{\vartheta_0}{|1-q|}\right),
\end{equation}
according to the periodicity of the defect, and 
\begin{equation}
r(0)=\hat{r}(\bar\vartheta)=\bar{r}\in(0,1),
\end{equation}
where $\tau$ is a parameter and $r(\tau)=\hat{r}(\vartheta(\tau))$,

The following solution
\begin{equation}
\label{eq:sol_diff_system}
\vartheta=\tau \ \hbox{mod}\left(\frac{\vartheta_0}{|1-q|}\right),\qquad
\hat{r}(\vartheta)=\dfrac{\bar{r}}{\sin w(\bar\vartheta)^{1/(q-1)}}\sin w(\vartheta)^{1/(q-1)}
\end{equation}
satisfies the system described above.

\section{Mathematical framework}\label{sec:model}

Drawing inspiration from biological shape formation and morphing, we present a mathematical framework that describes the shape deformations of an LCN with imprinted director field, $\n_0$ and order parameter $s$. We assume that an external stimulus like heat or light has been applied to our system, causing the order parameter to change from $s_0$ to $s$. Our framework explains how deformations occur to relieve the mechanical stress induced by the nematic director and the change in the order parameter.  

Let $\body_0$ be the reference configuration of the LCN under consideration. Our model is based on the following assumptions \cite{Cirak:2014:CLBW,Mihai:2020b:MG}:
\begin{itemize}
\item[(a)] The degrees of orientation $s_0$ and $s$ are \emph{prescribed} functions on $\body_0$. The LCN is aligned in the nematic phase according to a specific function $s$ and director field $\n_0$, as a result of an external stimulus that has altered the nematic order from $s_0$ to $s$. For $s_0=0$, the sample was in the isotropic phase; for example, in the context of morphogenesis, this setup can be understood as if cells, at that moment, adopt a specific degree of orientation dictated by $s$, forming a defect at locations where $s$ vanishes. This is an active impulse that is yet to be understood, but it leads to a known function $s$ and orientation field $\n_0$.
\item[(b)] All points of $\body_0$ suffer a deformation described by
\begin{equation}
\label{eq:phi_def}
\bm\varphi:\quad
\begin{aligned}
\body_0&\to\body=\bm\varphi\left(\body_0\right), \\
\bm X&\to\bm x=\bm\varphi(\bm X),
\end{aligned}
\end{equation}
where $\body$ is the current (deformed) configuration of the LCN.
\item[(c)] Since $\n_0$ is constrained to follow the elastic deformation, $\n_0\in T_{\bm X}\body$, the tangent space to $\body_0$ at $\bm X$, is mapped into $\n\in T_{\bm x}\body$, the tangent space to $\body$ at $\bm x$, hence 
\begin{equation}
\label{eq:n_deformed}
\n(\bm X)=\frac{\bm{\mathrm{F}}\n_0}{|\bm{\mathrm{F}}\n_0|},
\end{equation}
where  $\bm{\mathrm{F}}=\nabla_{\bm X}\bm\varphi$ is the \emph{gradient of the macroscopic deformation}.
\item[(d)] The material is incompressible, hence
\begin{equation}
\label{eq:incompressibility}
\det\bm{\mathrm{F}}=1.
\end{equation}
\end{itemize}

\subsection{Free-energy functional}

We adopt a variational approach to define an energy associated with the shape deformations of an LCN, and derive general equilibrium and stability conditions by considering the first and second variations of the energy functional. The energy associated with these deformations consists of two contributions,
\begin{equation}
\label{eq:free_energy}
\free[\bm\varphi]=\free_{\mathrm{N}}[\bm\varphi]+\free_{\mathrm{EL}}[\bm\varphi]\nonumber.
\end{equation}
In our model, the macroscopic deformation is due to the material’s response to two distinct mechanisms: on one hand, the mechanical stress induced by the distortion of the director field in the reference configuration, and on the other hand, changes in the nematic order parameter $s$. The first cause of activation is captured by the nematic contribution to the free energy $\free_{\mathrm{N}}[\bm\varphi]$, which retains no memory of $s_0$. Indeed, $\free_\mathrm{N}$ quantifies the energetic cost associated with the distortion of the director field $\n$ in the deformed configuration, and accounts only for the order parameter $s$ and $\n$ in $\body$. The elastomer contribution $\free_{\mathrm{EL}}[\bm\varphi]$ accounts for how variations in the degree of orientation can drive the system out of equilibrium, leading to shape changes. 

The first contribution $\free_{\mathrm{N}}[\bm\varphi]$ is associated with the nematic nature of the LCN, and accounts for spatial variations in the director field $\n$. In the one-constant approximation, it takes the form
\begin{equation}
\label{eq:free_nematic}
\free_{\mathrm{N}}[\bm\varphi]=\int_\body k_\mathrm{E} s(\bm\varphi^{-1}(\bm x))^2\left|\nabla_{\bm x}\n(\bm\varphi^{-1}(\bm x))\right|^2\dd\bm x,
\end{equation}
where $k_\mathrm{E}>0$ is an elastic constant characteristic of the material. It measures the cost associated with producing a distortion from the natural state, which, for nematic systems, corresponds to any uniform director field. $\free_{\mathrm{N}}$ is frame-indifferent and it is even in $\n$, since $\n$ is non-polar. This quantity is defined on the deformed configuration, and depends only on the gradient of the current director $\n$, and order parameter $s$. We recall that both $s$ and $\n$ were expressed in terms of $\bm X\in\body_0$; specifically, $s$ is a prescribed function on $\body_0$, while $\n$ is determined by the deformation through equation \eqref{eq:n_deformed}.

The formulation \eqref{eq:free_nematic} follows Ericksen's continuum theory for nematic LCs \cite{ericksen:liquid}, which aims to provide a comprehensive treatment of defects with arbitrary space dimension. When $s$ vanishes, molecular orientation is completely disordered, indicating a transition to the isotropic phase at that point of $\body$ without a temperature change; in this framework, defects correspond to localized isotropic regions. While the classical theory, with a constant $s$, adequately describes point defects in three-dimensional ($3D$) geometries, it fails in the case of line and plane defects where an infinite energy is obtained. Allowing the degree of orientation $s$ to vary in space mitigates singularities in $\n$, enabling the LC to locally transition to the isotropic phase wherever the classical theory predicts singularities in $\n$ with infinite energy. To account for variations in $s$, an additional contribution proportional to $|\nabla_{\bm X}s(\bm X)|^2$ is envisioned by Ericksen's theory. However, since $s$ is prescribed on $\body_0$, this term contributes only as an additional constant in our formulation.

The second contribution $\free_{\mathrm{EL}}$, describes the elastic response of the polymer network to deformations, taking into account the anisotropic distribution of monomers before and after the change in the order parameter, from $s_0$ to $s$. It is therefore based on the well-known phenomenological \emph{neoclassical strain-energy function} describing liquid crystal elastomers (LCEs) \cite{bladon:deformation,warner1988:theory,warner1991:elasticity}. Under our hypothesis $(a)$, which asserts that $s_0$ and $s$ are prescribed function of $\body_0$, it takes the generic form \cite{singh2023:bending,sonnet2022:photoresponsive}
\begin{equation}
\label{eq:neo_classical}
\free_{\mathrm{EL}}=\frac{\mu_0}{2}\int_{\body_0}\tr\left(\bF^\mathrm{T}\bL^{-1}\bF\bL_0\right)\dd \bm X,
\end{equation}
where $\bF$ represents the deformation gradient, $\mu_0$ is the shear elastic modulus at infinitesimal strain, and the tensors $\bL$ and $\bL_0$ are step-length tensors corresponding to the reference and current configurations, respectively \cite{Warner:2007:WT}. The reference configuration $\body_0$ for this contribution is described by $s_0$ and $\n_0$, which represent the scalar nematic order and the nematic director prior to stimulation (by heating or illumination say), while $s$ and $\n$ represent the scalar order and the nematic director in the present (activated) configuration $\body$. The distortion of $\n$ in $\body$ is instead captured by the nematic contribution to the free energy, $\free_{\mathrm{N}}$.

We represent $\bL_0$ and $\bL$ as \cite{Mihai:2020b:MG}
\begin{equation}
\label{eq:step_tensors}
\bL_0=a_0^{-1/3}\left[\left(a_0-1\right)\n_0\otimes\n_0+\bm{\mathrm{I}}\right], \qquad \bL=a^{-1/3}\left[\left(a-1\right)\n\otimes\n+\bm{\mathrm{I}}\right].
\end{equation}
In \eqref{eq:step_tensors}, $a_0>1$ and $a>1$ depend on $s_0$ and $s$, respectively. In the case of activation by light or temperature for example, $a$ also depends on how the light or temperature interacts with the material. 

We recall that, according to our model, $s_0$ and $s$ span $[0,1]$, and are prescribed functions on $\body_0$. Therefore, $a_0=a_0(s_0)$ and $a=a(s)$ are given on $\body_0$. When $a=a_0\equiv 1$, $\bL_0$ and $\bL$ in \eqref{eq:step_tensors} reduce to the identity tensor and the energy function \eqref{eq:neo_classical} reduces to the classical \emph{neo-Hookean formula} of isotropic rubber elasticity \cite{Treloar:1944}.

{\color{black} In the setting we are interested in, even though the neoclassical strain energy function in \eqref{eq:neo_classical}, which involves the director fields $\n_0$ and $\n$ and the order parameters $s_0$ and $s$, plays a role in determining the ground state of the elastic energy, it does not account for the spatial variation of the director field, which is measured by its gradient. In the presence of a disclination, for example, the director field becomes highly distorted, and the nematic contribution to the free energy in \eqref{eq:free_nematic} tends to diverge in three dimensions near the defect core. Therefore, when the nematic phase is well established away from the defect (i.e., when $s$ is sufficiently different from zero, typically between $0.3$ and $0.7$ \cite{brown1973:structure}), this contribution cannot be neglected.}

Before building upon \eqref{eq:free_nematic} and \eqref{eq:neo_classical} the free-energy functional that we shall study further, we find it useful to rescale all lengths according to the characteristic length scale set by the diameter of $\body_0$, denoted as $\diam$. After rewriting the nematic contribution in \eqref{eq:free_nematic}, originally expressed as an integral over the deformed configuration, in terms of the reference configuration using \eqref{eq:phi_def} and the incompressibility condition \eqref{eq:incompressibility}, {\color{black} we rescale the position vector $\bm X$ by $\diam$ to make it dimensionless}, keeping its name unaltered. We then denote the rescaled initial domain by ${\mathcal{\bar{B}}_0}$, and {\color{black} by defining the dimensionless elastic constant}
\begin{equation}
\label{eq:k_s_def}
\kappa=\dfrac{2k_\mathrm{E}}{\mu_0\ \diam^2},
\end{equation}
we arrive at the following reduced functional, which is an appropriate dimensionless form of $\free[\bm\varphi]$:
\begin{equation}
\label{eq:free_rescaled}
\mathcal{F}[\bm\varphi]=\frac{2\free[\bm\varphi]}{\mu_0 \ \diam^3}=\int_{\mathcal{\bar{B}}_0}\left[\kappa s(\bm X)^2\left|\nabla_{\bm X}\n\bm{\mathrm{F}}^{-1}\right|^2 +\tr\left(\bF^\mathrm{T}\bL^{-1}\bF\bL_0\right)\right]\dd\bm X.
\end{equation}
{\color{black}Since $[k_\mathrm{E}]=\mathrm{N}=\mathrm{J}/\mathrm{m}$, $[\mu_0]=\mathrm{Pa}=\mathrm{N}/\mathrm{m}^2=\mathrm{J}/\mathrm{m}^3$, and $[\free]=\mathrm{J}$, all terms in \eqref{eq:k_s_def} and \eqref{eq:free_rescaled} are properly normalized to ensure the energy is expressed in a dimensionless form.}

The constant $\kappa$ defined in \eqref{eq:k_s_def} is a measure of the balance between the nematic elasticity, encapsulated in the elastic constant $k_{\mathrm{E}}$, and the material stiffness, described by $\mu_0$. It is the interplay between nematic elasticity, which seeks to minimize distortions in the director field, and mechanical stiffness, which resists deformation, that determines the resulting morphology. 

\subsection{Equilibrium and stability conditions}

We derive equilibrium and stability conditions by considering the first and second variations of $\mathcal{F}[\bm\varphi]$ in \eqref{eq:free_rescaled}, with the detailed computations deferred to Appendix~\ref{sec:first_second_var_app}. In the following, the first variation of $\mathcal{F}$ is denoted by $\delta_1\mathcal{F}$ and is associated with a test field $\bG_1=\delta_1\bF$ that represents an admissible variation of the deformation gradient $\bF$ that satisfies the incompressibility assumption (d). As shown in the appendix, $\bG_1$ is accordingly subject to the condition
\begin{equation}
\label{eq:perturb_admissible}
\bG_1\cdot\bF^{-\mathrm{T}}=0.
\end{equation}
The second variation of $\mathcal{F}$ is instead denoted as $\delta_2\delta_1\mathcal{F}$, and is obtained by introducing another admissible variation $\bG_2=\delta_2\bF$ of $\bF$, such that $\bG_2\cdot\bF^{-\mathrm{T}}=0$, and further perturbing $\delta_1\mathcal{F}$.

Making use of the identities recorded in Appendix~\ref{sec:first_second_var_app}, the first variation of the dimensionless energy functional $\mathcal{F}$ at the field $\bm\varphi$ results in a linear functional of $\bG_1$ given by
\begin{align}
\label{eq:equilibrium_condition}
\delta_1\mathcal{F}[\bm\varphi](\bG_1)=\int_{\mathcal{B}_0}&\Big[2\kappa s^2\bm{\mathrm{A}}[\vv_1,\bm{\mathrm{G}}_1]\cdot(\nabla\n)\bm{\mathrm{F}}^{-1}+\tr\left(\bG_1^{\mathrm{T}}\bL^{-1}\bF\bL_0\right)+\tr\left(\bF^{\mathrm{T}}\bL^{-1}\bG_1\bL_0\right)\nonumber\\
&+a^{1/3}\left(a^{-1}-1\right)\tr\left(\bF^{\mathrm{T}}(\n\otimes\vv_1+\vv_1\otimes\n)\bF\bL_0\right)\Big]\dd \bm X=0,
\end{align}
where
\begin{equation}
\label{eq:A_tens_def}
\bm{\mathrm{A}}[\vv_1,\bG_1]=(\nabla\vv_1)\bm{\mathrm{F}}^{-1}-(\nabla\n)\bm{\mathrm{F}}^{-1}\bm{\mathrm{G}}_1\bm{\mathrm{F}}^{-1},
\end{equation}
with $\vv_1$ representing the variation of $\n$ in \eqref{eq:n_deformed}, which is obtained as
\begin{equation}
\label{eq:vv_1_def_main}
\vv_1=\delta_1\n=\dfrac{1}{|\bF\n_0|}\left(\bm{\mathrm{I}}-\n\otimes\n\right)\bm{\mathrm{G}_1}\n_0.
\end{equation}
The equilibrium conditions for our system are given by requiring that
\begin{equation}
\label{eq:equilibrium_equation}
\delta_1\mathcal{F}[\bm\varphi](\bG_1)=0
\end{equation}
for any admissible test field $\bG_1$ such that \eqref{eq:perturb_admissible} holds.

Similarly, the second variation of $\mathcal{F}$ at the field $\bm\varphi$ is obtained in Appendix~\ref{sec:first_second_var_app} as the following bilinear form in the admissible perturbations $\bG_1$ and $\bG_2$ of $\bF$,
\begin{align}
\label{eq:stability_condition}
\delta_2\delta_1\mathcal{F}&[\bm{\varphi}](\bG_1,\bG_2)=\int_{\mathcal{B}_0}\left\{2\kappa s^2\left[\left((\nabla\bm\xi)\bF^{-1}-\bm{\mathrm{A}}[\vv_1,\bm{\mathrm{G}}_1]\bm{\mathrm{G}}_2\bF^{-1}-\bm{\mathrm{A}}[\vv_2,\bm{\mathrm{G}}_2]\bm{\mathrm{G}_1}\bF^{-1}\right)\cdot(\nabla\n)\bF^{-1}\right.\right.\nonumber\\
&\left.\left.+\bm{\mathrm{A}}[\vv_1,\bm{\mathrm{G}}_1]\cdot\bm{\mathrm{A}}[\vv_2,\bm{\mathrm{G}}_2]\right]+\tr\left(\bG_1^{\mathrm{T}}\bL^{-1}\bG_2\bL_0\right)+\tr\left(\bG_2^{\mathrm{T}}\bL^{-1}\bG_1\bL_0\right)\right.\nonumber\\
&\left.+a^{1/3}\left(a^{-1}-1\right)\left[\tr\left(\bG_1^{\mathrm{T}}(\n\otimes\vv_2+\vv_2\otimes\n)\bF\bL_0\right)+\tr\left(\bG_2^{\mathrm{T}}(\n\otimes\vv_1+\vv_1\otimes\n)\bF\bL_0\right)\right.\right.\nonumber\\
&\left.\left.+\tr\left(\bF^{\mathrm{T}}(\n\otimes\vv_2+\vv_2\otimes\n)\bG_1\bL_0\right)+\tr\left(\bF^{\mathrm{T}}(\n\otimes\vv_1+\vv_1\otimes\n)\bG_2\bL_0\right)\right.\right.\nonumber\\
&\left.\left.+\tr\left(\bF^{\mathrm{T}}(\vv_2\otimes\vv_1+\vv_1\otimes\vv_2)\bF\bL_0\right)+\tr\left(\bF^{\mathrm{T}}(\n\otimes\xi+\xi\otimes\n)\bF\bL_0\right)\right]\right\}\dd\bm X.
\end{align}
Here, 
\begin{equation}
\label{eq:A_tens_def_2}
\bm{\mathrm{A}}[\vv_2,\bG_2]=(\nabla\vv_2)\bm{\mathrm{F}}^{-1}-(\nabla\n)\bm{\mathrm{F}}^{-1}\bm{\mathrm{G}}_2\bm{\mathrm{F}}^{-1},
\end{equation}
while $\vv_2$ and $\bm\xi$ represent the variations of $\n$ in \eqref{eq:n_deformed} and $\vv_1$ in \eqref{eq:vv_1_def_main}, respectively, through $\bG_2$, and are given by
\begin{subequations}
\label{eq:stab_def}
\begin{align}
\vv_2&=\dfrac{1}{|\bm{\mathrm{F}}\n_0|}\left(\bm{\mathrm{I}}-\n\otimes\n\right)\bm{\mathrm{G}}_2\n_0,\\
\bm\xi&=-\frac{1}{|\bF\n_0|}\left[(\vv_1\otimes\n)\bG_2\n_0+(\vv_2\otimes\n+\n\otimes\vv_2)\bG_2\n_0\right].
\end{align}
\end{subequations}

The stability of a specific equilibrium configuration, given by a solution $\hat{\bm{\varphi}}$ of the equilibrium equations, is equivalent to requiring that 
\begin{equation}
\label{eq:sec_var_positive}
\delta_1\delta_2\mathcal{F}[\hat{\bm{\varphi}}](\bG_1,\bG_2)>0
\end{equation}
for all pairs $(\bG_1,\bG_2)$ of admissible  non-vanishing perturbations of $\hat{\bF}=\nabla\hat{\bm\varphi}$ that, consistent with \eqref{eq:perturb_admissible}, satisfy $\bm{\mathrm{G}}_1\cdot\hat{\bm{\mathrm{F}}}^{-\mathrm{T}}=0$ and $\bm{\mathrm{G}}_2\cdot\hat{\bm{\mathrm{F}}}^{-\mathrm{T}}=0$.

\section{Out-of-plane perturbations}\label{sec:examples}

Using our mathematical framework, we aim to describe the \emph{out-of-plane shape deformations} of an initial LCN sheet represented in the $3D$ space as a flat slab of thickness $H$ around a disk-shaped mid-surface $\mathcal{S}_0$ of radius $R$,
\begin{equation}
\label{eq:flat_LCN}
\body_0=\mathcal{S}_0\times\left[-\frac{H}{2},\frac{H}{2}\right].
\end{equation}
A central defect $\m_0$ of topological charge $q$, described by \eqref{eq:n_0_q}, is imprinted on $\mathcal{S}_0$, and extended \emph{uniformly} across the cross-section.
Following a well-established practice (see, e.g., [\cite{degennes:physics}, p.~171]) we posit that the energy concentration near defects causes a localized transition to the isotropic phase, which constitutes a defect core. The energy associated with such a phase transition is proportional to the core’s area and will be taken as approximately fixed. Letting $r_{\mathrm{c}}$ denote the core’s size, we set $r_{\mathrm{c}}=\varepsilon R$, where a sensible value for the parameter $\varepsilon$ is $\varepsilon\approx 10^{-3}$, and we prescribe the degrees of order $s_0$ and $s$ on $\body_0$ to be zero around the defect, and constant throughout the remaining region, meaning that the nematic phase is well established away from the defect. More precisely, 
\begin{equation}
\label{eq:s_out_of_plane}
  s_0= \begin{cases}
      0 & \, \hbox{on} \qquad \mathbb{B}_{r_{\mathrm{c}}}(\bm 0)\\
      \bar{s}_0 & \, \hbox{on} \quad \body_0\setminus\mathbb{B}_{r_{\mathrm{c}}}(\bm 0) 
    \end{cases}\, \qquad  s= \begin{cases}
      0 & \, \hbox{on} \quad \mathbb{B}_{r_{\mathrm{c}}}(\bm 0)\\
      \bar{s} & \, \hbox{on} \quad \body_0\setminus\mathbb{B}_{r_{\mathrm{c}}}(\bm 0) 
    \end{cases}\,.
\end{equation} 
where $\mathbb{B}_{r_{\mathrm{c}}}(\bm 0)$ is a cylinder of radius $r_{\mathrm{c}}$ around the disclination. Accordingly, the values $\bar a_0=a_0(\bar s_0)$ and $\bar a=a(\bar s)$ are prescribed on $\body_0\setminus\mathbb{B}_{r_{\mathrm{c}}}(\bm 0)$.

Since we are interested in the stability of the LCN, our base deformation corresponds to the identity vector, and so
\begin{equation}
\label{eq:phi_flat}
\varphi_0(\bm X)=\bm X=r\e_r+z\e_z,
\end{equation}
and so the deformation gradient coincides with the identity tensor, $\bm{\mathrm{F}}_0=\bm{\mathrm{I}}$, while $\n=\m_0$ by \eqref{eq:n_deformed}. For out-of-plane deformations of the flat configuration, we refer to the experimental study in \cite{McConney:2013:etal}, where 3D photo-activable LCNs, prepared with an imprinted central defect of topological charge $q$, which is different for each sample, deform and adopt a complex topography specific to that defect.

Our model assumes that $\bar s_0$ and $\bar{s}$ in \eqref{eq:s_out_of_plane} represent the order parameters of the sample before and after the activation, respectively, of the LCN sample. For example, in the case of \cite{McConney:2013:etal}, the value $\bar s$ is lower than the one at which the sample was originally prepared, $\bar s_0$. This reduction in the degree of order, along with the mechanical stress induced by $\m_0$, plays a role in the stability of the flat LCN disk.

We consider a relatively simple case involving only out-of-plane perturbations of the flat configuration  $\bm\varphi_0$, i.e., along the $z$-direction. These perturbations $\delta_i\bm\varphi$ are described by generic functions $\delta h_i$, which depends on both the radial and polar coordinates $(r,\vartheta)$, as
\begin{equation}
\label{eq:out_of_plane_perturbations}
\delta_i\bm\varphi_0=\delta h_i(r,\vartheta)\e_z,
\end{equation}
where $i=1,2$ in accordance with the nomenclature used for the first and second variations of $\mathcal{F}$ in the previous section.

Thus the admissible variations $\bG_i$ of $\bF_0$, result to be defined as
\begin{equation}
\label{eq:variation_out_of_plane}
\bm{\mathrm{G}}_i=\delta_i\bm{\mathrm{F}}_0=\nabla\delta_i\bm\varphi_0=\e_z\otimes\nabla\delta h_i=\frac{\partial \delta h_i}{\partial r}\e_z\otimes\e_r+\frac{1}{r}\frac{\partial \delta h_i}{\partial \vartheta}\e_z\otimes\e_\vartheta,
\end{equation}
where $\nabla=\partial_r+\frac{1}{r}\partial_\vartheta+\partial_z$ is the gradient in polar coordinates.

We note that each $\bG_i$ in \eqref{eq:variation_out_of_plane} is consistent with the incompressibility constraint, as it satisfies \eqref{eq:perturb_admissible}, which, in this case, reduces to $\tr\bG_i=0$.

By \eqref{eq:variation_out_of_plane}, we find in Appendix~\ref{sec:out_of_plane_app} that the flat disk $\bm\varphi_0$ with the imprinted director field $\m_0$ in \eqref{eq:n_0_q} is an equilibrium configuration for every $q$, while $\bar a_0, \, \bar a>1$ and $\kappa>0$. It should be observed that in the absence of a tendency for monomers in polymer chains to distribute anisotropically, that is, when $\kappa=0$ and $\bar a_0=\bar a\equiv1$ and the material consists of an isotropic rubber, the flat configuration $\bm\varphi_0$ corresponds to the elastic ground state of the system. 

To analyze the stability of the flat configuration for $\kappa>0$, $\bar a_0>1$, and $\bar a>1$, we will study the second variation \eqref{eq:stability_condition} of the functional $\mathcal{F}$ at $\bm\varphi_0$. In Appendix~\ref{sec:out_of_plane_app}, we show that it reduces to
\begin{align}
\label{eq:second_variation_out_of_plane}
&\delta_2\delta_1\mathcal{F}[\bm\varphi_0](\bm{\mathrm{G}_1},\bm{\mathrm{G}}_2)=\nonumber\\
&=\frac{H}{R}\left(\frac{\bar a}{\bar{a}_0}\right)^{1/3}\int_\varepsilon^1\int_0^{2\pi}\left[\left(-\frac{k q^2}{r}+r\gamma\right)v_1v_2+kr\nabla v_1\cdot\nabla v_2+r\delta h_{1,r}\delta h_{2,r}+\frac{1}{r}\delta h_{1,\vartheta}\delta h_{2,\vartheta}\right]\dd r \dd \vartheta,
\end{align}
where
\begin{equation}
\label{eq:k_gamma_elastomeric}
k=\frac{\kappa\bar s^2 \bar a_0^{1/3}}{\bar a^{1/3}}, \qquad \gamma=\frac{\bar a_0+1-2\bar a}{\bar a},
\end{equation}
and
\begin{equation}
\label{eq:v_1_flat}
v_i=\delta h_{i,r}\cos w+\frac{1}{r}\delta h_{i,\vartheta}\sin w,
\end{equation}
$w=w(\vartheta)$ defined in \eqref{eq:n_0_q}. Here, all lengths are rescaled to $R$ instead of $\diam$, and $\nabla=\partial_r+\frac{1}{r}\partial_\vartheta+\partial_z$ represents the gradient in polar coordinates. Since $a_0$ and $a$ are both greater than $1$, it follows that $\gamma$ in \eqref{eq:k_gamma_elastomeric} satisfies
\begin{equation}
\label{eq:gamma_bound}
\gamma>-2.
\end{equation}

Note that two parameters, $k$ and $\gamma$ defined in \eqref{eq:k_gamma_elastomeric}, appear in the second variation. The parameter $\gamma$ represents the tendency of monomers in the polymer chains to distribute anisotropically in response to an external stimulus (such as heat or light), and it vanishes for isotropic rubber where $a=a_0=1$. Its value changes according to variations in the degree of order, from $\bar s_0$ to $\bar s$, and thus depends on the values of $\bar a_0$ and $\bar a$. For this reason, we consider $\gamma$ as a known quantity based on the experimental setup. The parameter $k$ can be interpreted as a reduced elastic constant that, according also to \eqref{eq:k_s_def}, measures the balance between the nematic elasticity, represented by $k_{\mathrm{E}}$, the material stiffness, represented by $\mu_0$, and the variation in $s$, represented by the ratio $\bar s^2 \bar a_0^{1/3}/\bar a^{1/3}$. 

For a given $\gamma>-2$, the onset of instability will be determined by the critical value of $k$ at which the flat configuration ceases to be a minimum for the dimensionless energy. Moreover, we will also characterize the nature of the corresponding buckling modes for which $\delta_1\delta_2\free[\bm\varphi_0](\bm{\mathrm{G}}_1,\bm{\mathrm{G}}_2)=0$.

\subsection{Deformation according to the topological charge}

To proceed, we apply a separation of variables to each function $\delta h_i$, expressing it as a product of a function $f_i$, depending on $r$, and a function $g_i$, depending $\vartheta$,
\begin{equation}
\label{eq:separation_of_variables}
\delta h_i(r,\vartheta)=f_i(r)g_i(\vartheta),
\end{equation}
and subject to the conditions
\begin{equation}
\label{eq:conditions_fg}
f_i(\varepsilon)=0, \quad g_i(0)=g_i(2\pi).
\end{equation}
The first condition in \eqref{eq:conditions_fg} is allowed by the translational invariance of the free energy $\mathcal{F}$ and states that $f_i$ vanishes around the defects, while the periodicity condition on $g_i$ ensures the continuity of $\delta h_i$.

Moreover, since we expect that the deformation of the flat configuration $\bm\varphi_0$ will have the same periodicity of the defect $\m_0$ imprinted in the LCN, we introduce an ansatz for the functions $g_i(\vartheta)$, given (up to a multiplicative constant) by
\begin{equation}
\label{eq:ansatz_g_i}
g_i(\vartheta)=g(\vartheta)=\begin{cases}
    |\sin w| & \text{for } q=\dfrac{2m+1}{2}, \, \, m\in\mathbb{Z}, \\
    \cos w & \text{for } q\in\mathbb{Z}.
  \end{cases}
  \end{equation}
Indeed, for a defect with topological charge $q$, $2|q-1|$ regions in which $\m_0$ and $-\m_0$ repeat periodically can be identified, and we expect that these regions of periodicity will be maintained by the deformation.

By substituting \eqref{eq:separation_of_variables} and \eqref{eq:ansatz_g_i} in \eqref{eq:second_variation_out_of_plane}, a localization argument detailed in Appendix~\ref{sec:solution_app} implies that, for a given $\gamma>-2$, $\delta_2\delta_1\mathcal{F}[\bm\varphi_0](\bG_1,\bG_2)$ vanishes for every $f_2$ satisfying \eqref{eq:conditions_fg} whenever there exists a critical value of the constant $k$ for which there exists at least one non-zero function $\bar f_1(r)$, defined up to a multiplicative constant, which satisfies the following equation on $(\varepsilon,1)$:
\begin{align}
\label{eq:bulk_eq}
&k r(A-A_w)f_1^{iv}+2k (A-A_w)f_1'''\nonumber\\
+&\frac{f_1''}{r}\left[\bar k \left((3q-2)A-(4q-3)A_w-B\right)-(A+\gamma(A-A_w))r^2\right]\nonumber\\
+&\frac{f_1'}{r^2}\left[-\bar k \left((3q-2)A-(4q-3)A_w-B\right)-(A+\gamma(A-A_w))r^2\right]\nonumber\\
+&\frac{f_1}{r^3}\left[\bar k \left(2q(q-1)(A-2A_w)+B(q-1)+C_w-2B_w\right)+(B+\gamma B_w)r^2\right]=0
\end{align}
with the boundary conditions
\begin{subequations}
\label{eq:conditions_f1_null_second_variation}
\begin{align}
&f_1(\varepsilon)=0\label{eq:bound_1},\\
&- k r(A-A_w)f_1'''(1)+\frac{f_1'(1)}{2}\left[ k \left(A(-5q+4)+A_w(6q-4)+2B\right)+2(A+\gamma(A-A_w))\right]\nonumber\\
&+\frac{f_1(1)}{2}\left[ k ((A-2A_w)(q-1)(2q-1)+2B_w(q-2)+B(q-1))+\gamma(2A_w-A)(q-1)\right]=0,\label{eq:bound_2}\\
&(A-A_w)f_1''(1)+\frac{(q-1)}{2}(A-2A_w)\left(f_1(1)-f'_1(1)\right)=0,\label{eq:bound_3}\\
&\varepsilon(A-A_w)f_1''(\varepsilon)+\frac{(q-1)}{2}(A-2A_w)\left(\frac{1}{\varepsilon}f_1(\varepsilon)-f'_1(\varepsilon)\right)=0.\label{eq:bound_4}
\end{align}
\end{subequations}
This critical value of $k$ provides an upper bound estimate of the threshold at which the flat configuration ceases to be a local minimum for the free energy, while the real part of the corresponding $\delta h_1=g(\vartheta)\bar f_1(r)$, with $g$ given by \eqref{eq:ansatz_g_i}, represents the associated bucking mode (defined up to an arbitrarily multiplicative constant). 

\begin{figure}[htbp]
    \centering
    \begin{subfigure}{0.45\textwidth}
        \centering
        \vspace{0.8cm}
        \includegraphics[width=\textwidth]{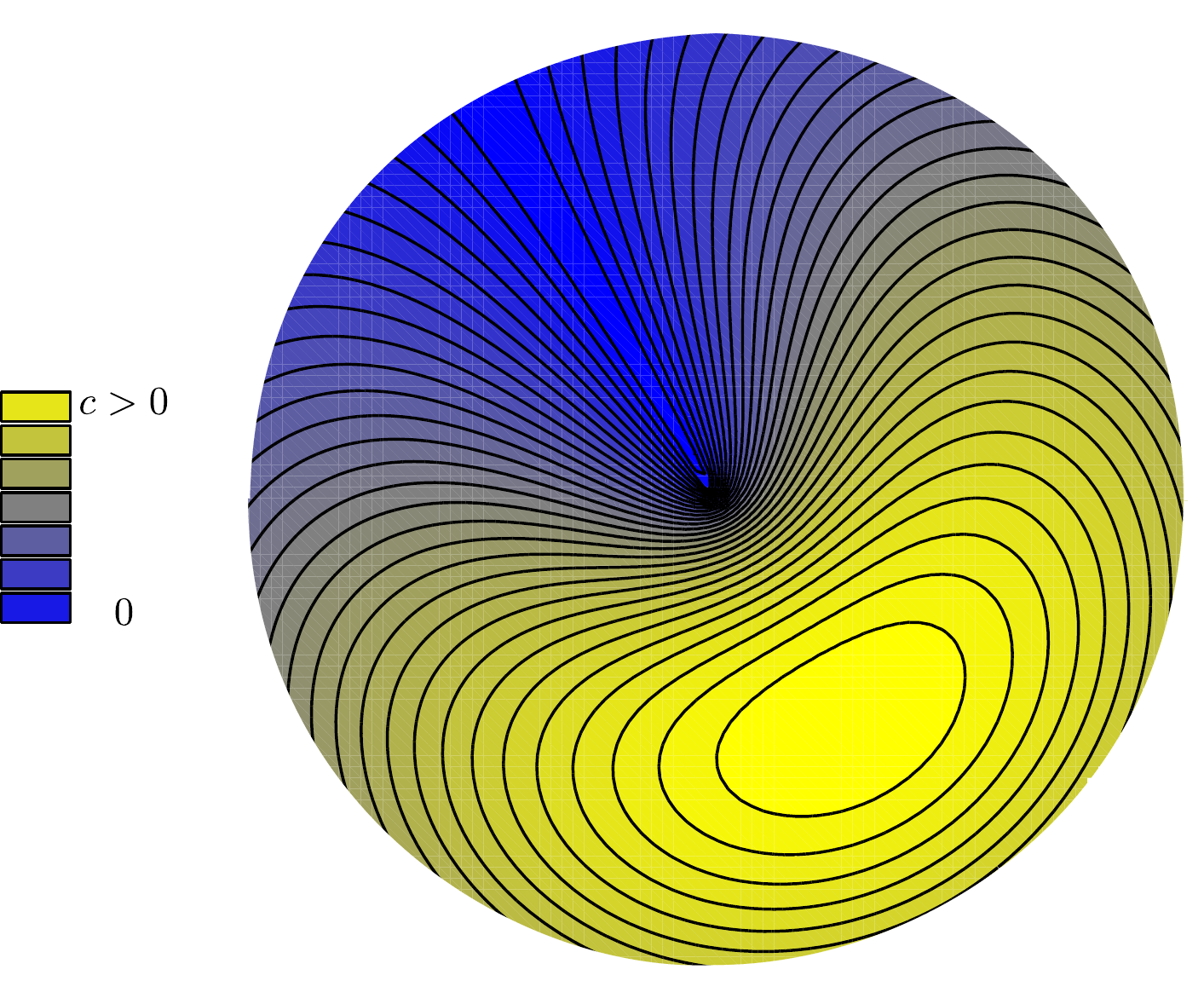}
        \caption{Contour plot of the buckling mode $\delta h_1$.}
        \label{fig:contour_1o2}
    \end{subfigure}  
    \hfill
    \begin{subfigure}{0.54\textwidth}
        \centering
        \includegraphics[width=0.4\textwidth]{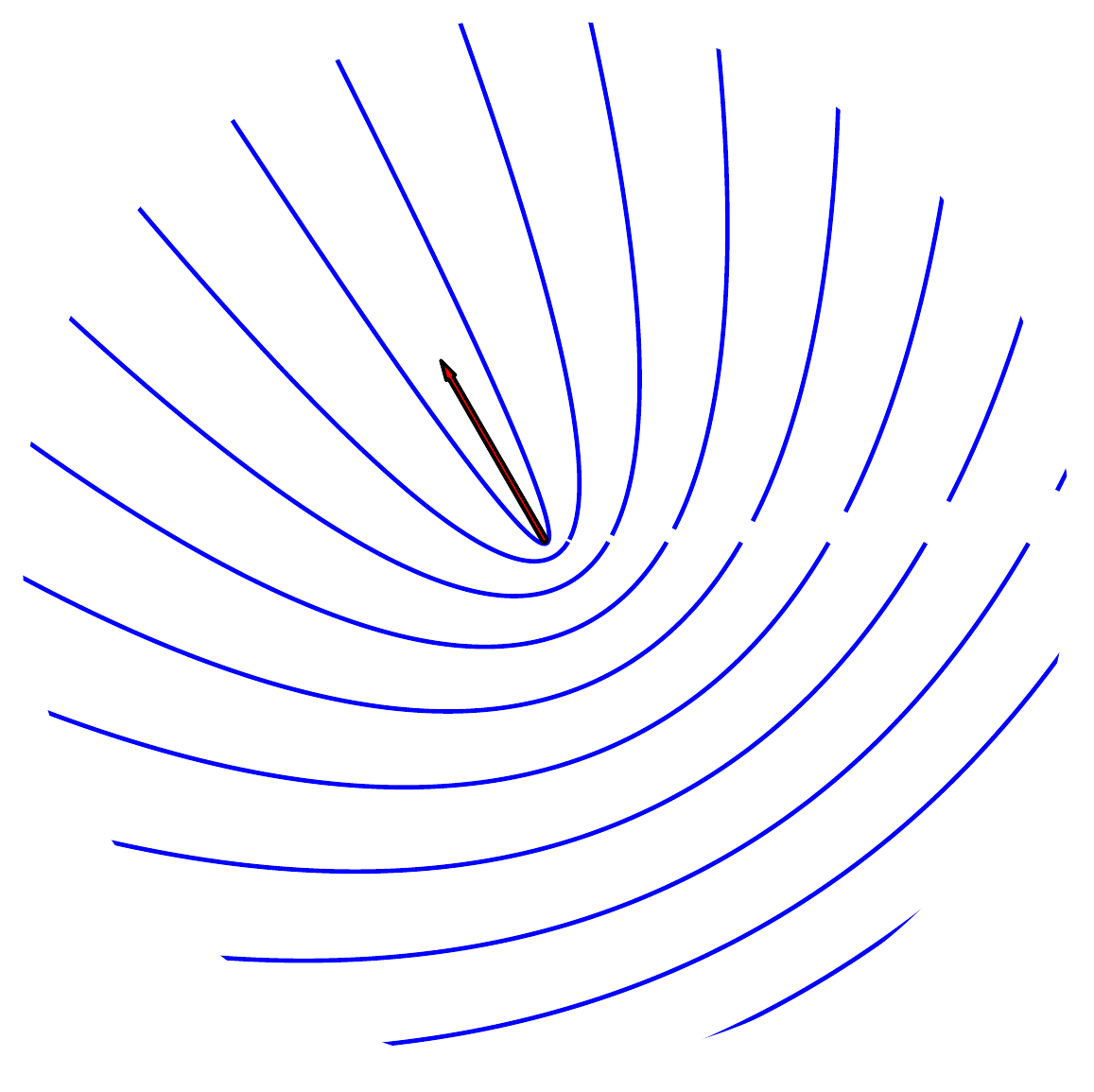}
        \caption{Field lines of $\m_0$ in equation \eqref{eq:n_0_q}.}
        \label{fig:lines_1o2}
        
         \vspace{0.5cm} 
        
        \includegraphics[width=\textwidth]{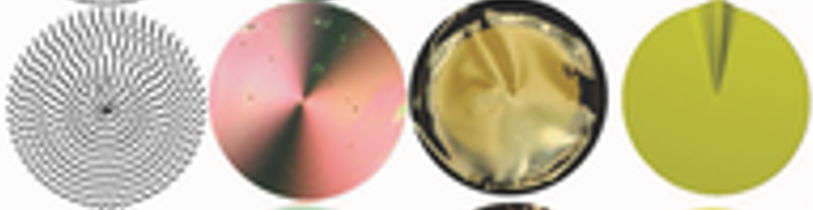}
        \caption{The $+1/2$ defect in LCN film \cite{McConney:2013:etal}.}
        \label{fig:delta_h_1o2}
    \end{subfigure}
    \caption{Illustration of the out-of plane buckling mode of deformation of LCN sheets with an imprinted director field $\m_0$ of topological charge $q=+\dfrac{1}{2}$ and $\vartheta_0=\dfrac{\pi}{3}$ in the reference configuration. For $\gamma=-1/2$, this occurs at the critical value $k\approx1.43$.}
    \label{fig:main_1o2}
\end{figure}

\begin{figure}[htbp]
    \centering
    \begin{subfigure}{0.45\textwidth}
        \centering  
        \vspace{0.95cm}
              \includegraphics[width=\textwidth]{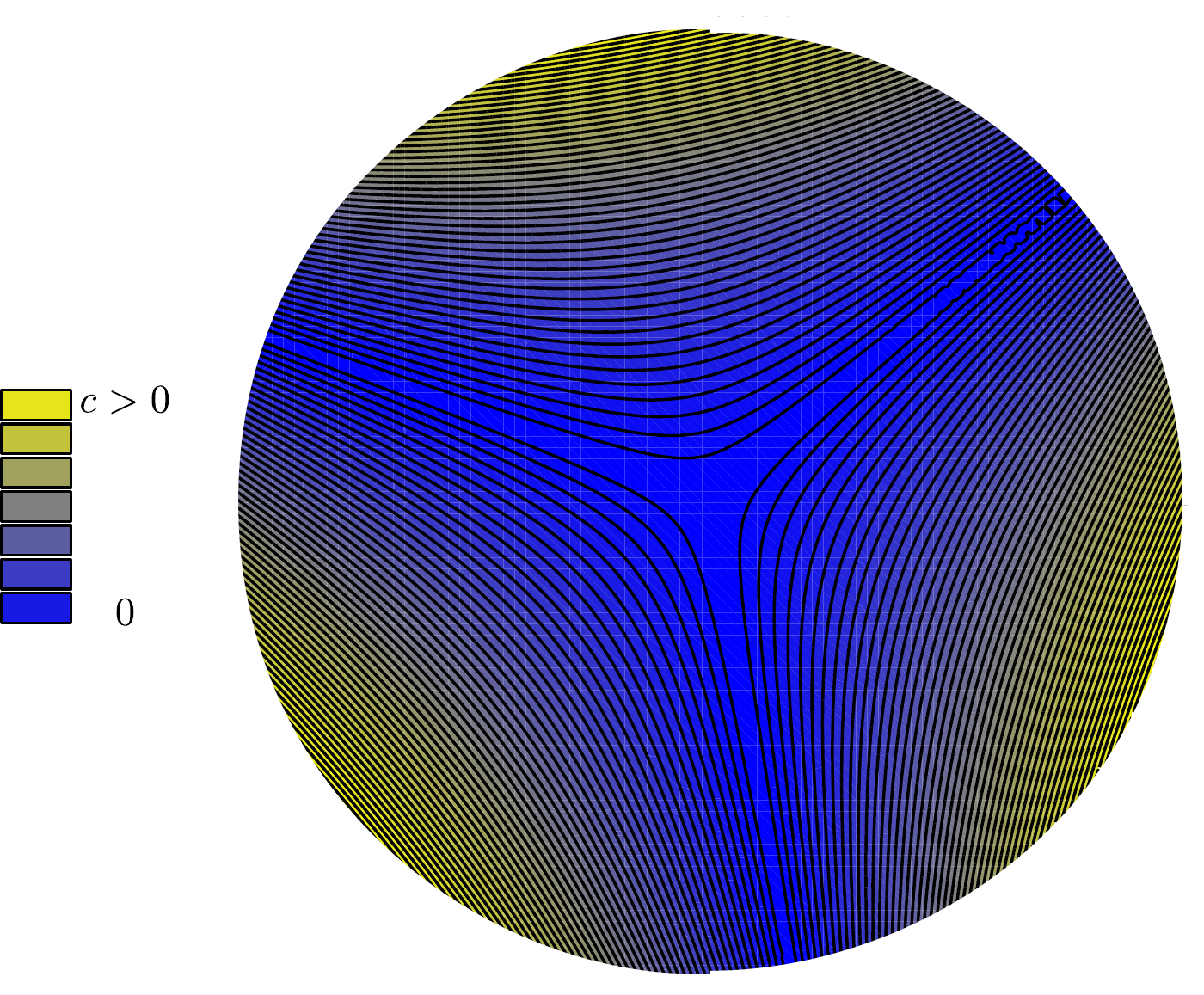}
        \caption{Contour plot of the buckling mode $\delta h_1$.}
        \label{fig:contour_minus_1o2}
    \end{subfigure}
    \hfill
    \begin{subfigure}{0.54\textwidth}
        \centering
        \includegraphics[width=0.4\textwidth]{defect_minus_1o2-eps-converted-to.pdf}
        \caption{Field lines of $\m_0$ in equation \eqref{eq:n_0_q}.}
        \label{fig:lines_minus1o2}
        
         \vspace{0.5cm} 
        
        \includegraphics[width=\textwidth]{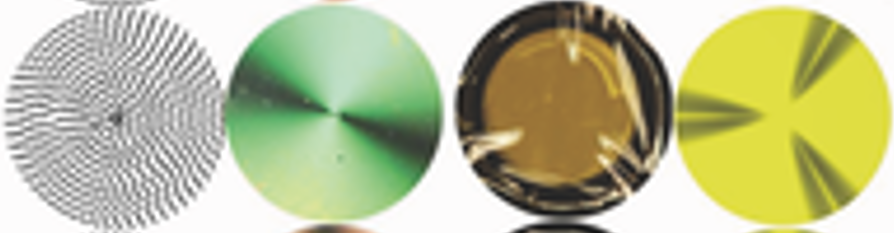}
        \caption{The $+1/2$ defect in LCN film \cite{McConney:2013:etal}.}
        \label{fig:delta_h_minus_1o2}
    \end{subfigure}
    \caption{Illustration of the out-of plane buckling mode of deformation of LCN sheets with an imprinted director field $\m_0$ of topological charge $q=-\dfrac{1}{2}$ and $\vartheta_0=\dfrac{\pi}{3}$ in the reference configuration. For $\gamma=-1/2$, this occurs at the critical value $k\approx1.02$.}
    \label{fig:main_minus_1o2}
\end{figure}

\begin{figure}[htbp]
    \centering
    \begin{subfigure}{0.45\textwidth}
        \centering
        \vspace{0.8cm}
        \includegraphics[width=\textwidth]{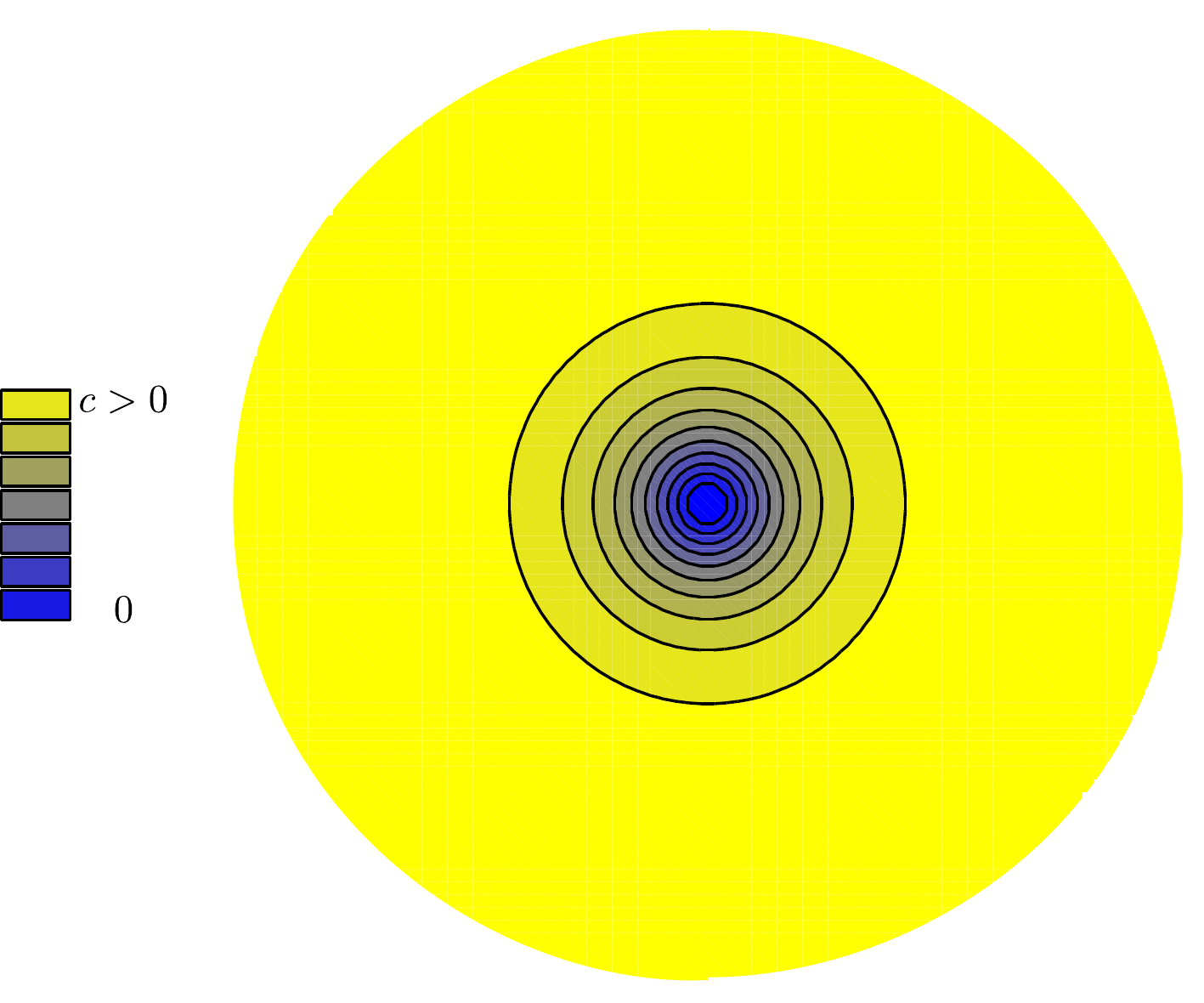}
        \caption{Contour plot of the buckling mode $\delta h_1$.}
        \label{fig:contour_1}
    \end{subfigure}
    \hfill
    \begin{subfigure}{0.54\textwidth}
        \centering
        \includegraphics[width=0.4\textwidth]{defect_plus_1-eps-converted-to.pdf}
        \caption{Field lines of $\m_0$ in equation \eqref{eq:n_0_q}.}
        \label{fig:lines_1}
        
         \vspace{0.5cm} 
        
        \includegraphics[width=\textwidth]{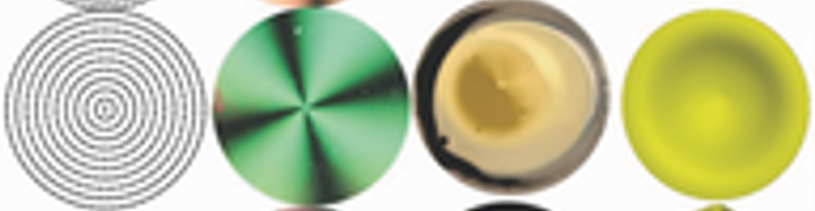}
        \caption{The $+1$ defect in LCN film \cite{McConney:2013:etal}.}
        \label{fig:delta_h_1}
    \end{subfigure}
    \caption{Illustration of the out-of plane buckling mode of deformation of LCN sheets with an imprinted director field $\m_0$ of topological charge $q=+1$ and $\vartheta_0=\dfrac{\pi}{3}$ in the reference configuration. For $\gamma=-1/2$, this occurs at the critical value $k\approx 0.09$.}
    \label{fig:main_1}
\end{figure}

\begin{figure}[htbp]
    \centering
    \begin{subfigure}{0.45\textwidth}
        \centering
        \vspace{0.8cm}
        \includegraphics[width=\textwidth]{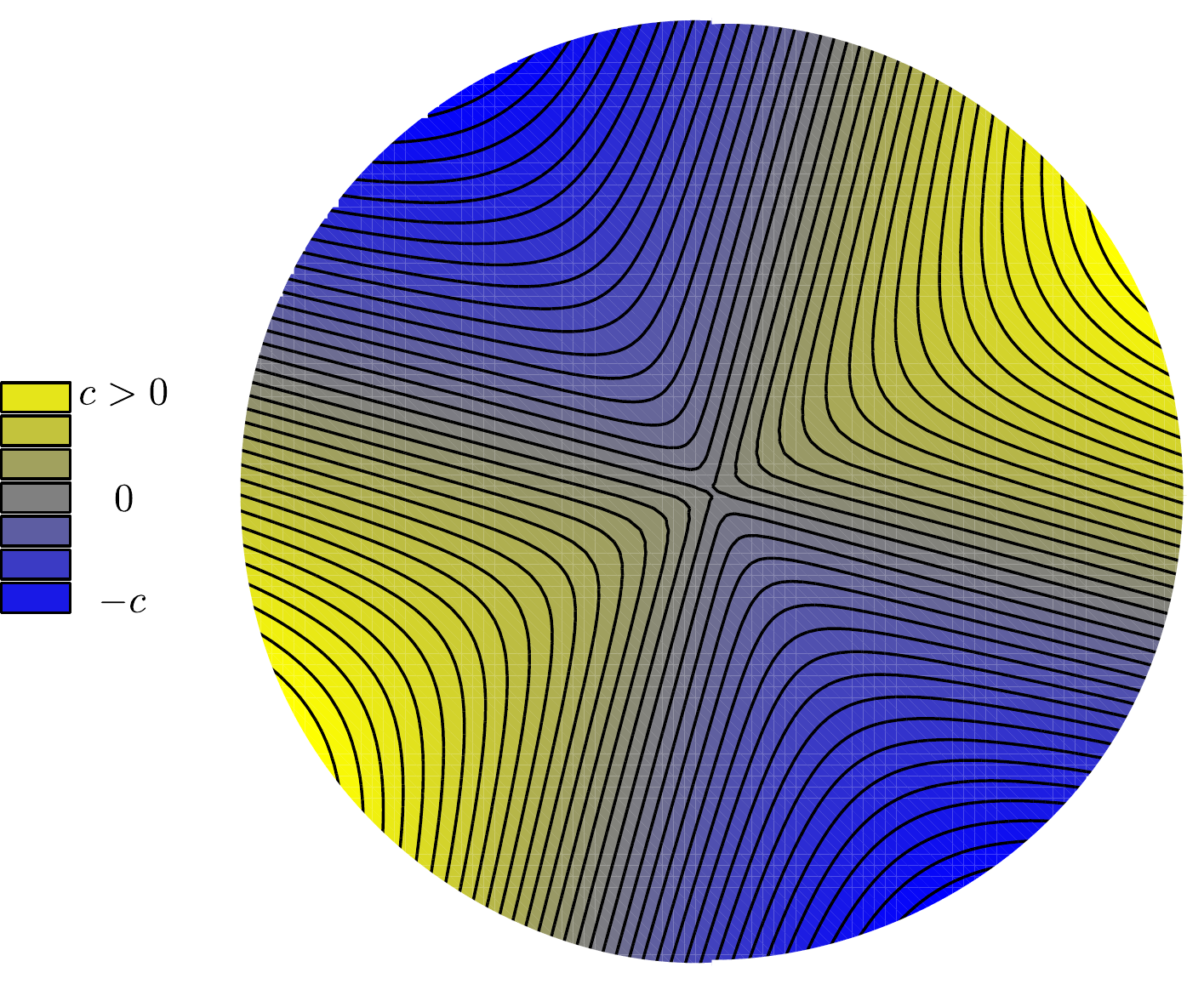}
        \caption{Contour plot of the buckling mode $\delta h_1$.}
        \label{fig:contour_minus_1}
    \end{subfigure}
    \hfill
    \begin{subfigure}{0.54\textwidth}
        \centering
        \includegraphics[width=0.4\textwidth]{defect_minus_1-eps-converted-to.pdf}
        \caption{Field lines of $\m_0$ in equation \eqref{eq:n_0_q}.}
        \label{fig:lines_minus_1}
        
         \vspace{0.5cm} 
        
        \includegraphics[width=\textwidth]{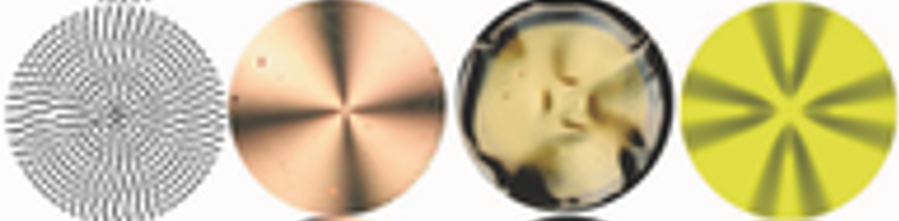}
        \caption{The $-1$ defect in LCN film \cite{McConney:2013:etal}.}
        \label{fig:delta_h_minus_1}
    \end{subfigure}
    \caption{Illustration of the out-of plane buckling mode of deformation of LCN sheets with an imprinted director field $\m_0$ of topological charge $q=-1$ and $\vartheta_0=\dfrac{\pi}{3}$ in the reference configuration. For $\gamma=-1/2$, this occurs at the critical value $k\approx 0.23$.}
    \label{fig:main_minus_1}
\end{figure}

\begin{figure}[htbp]
    \centering
    \begin{subfigure}{0.45\textwidth}
        \centering
        \vspace{1.15cm}
        \includegraphics[width=\textwidth]{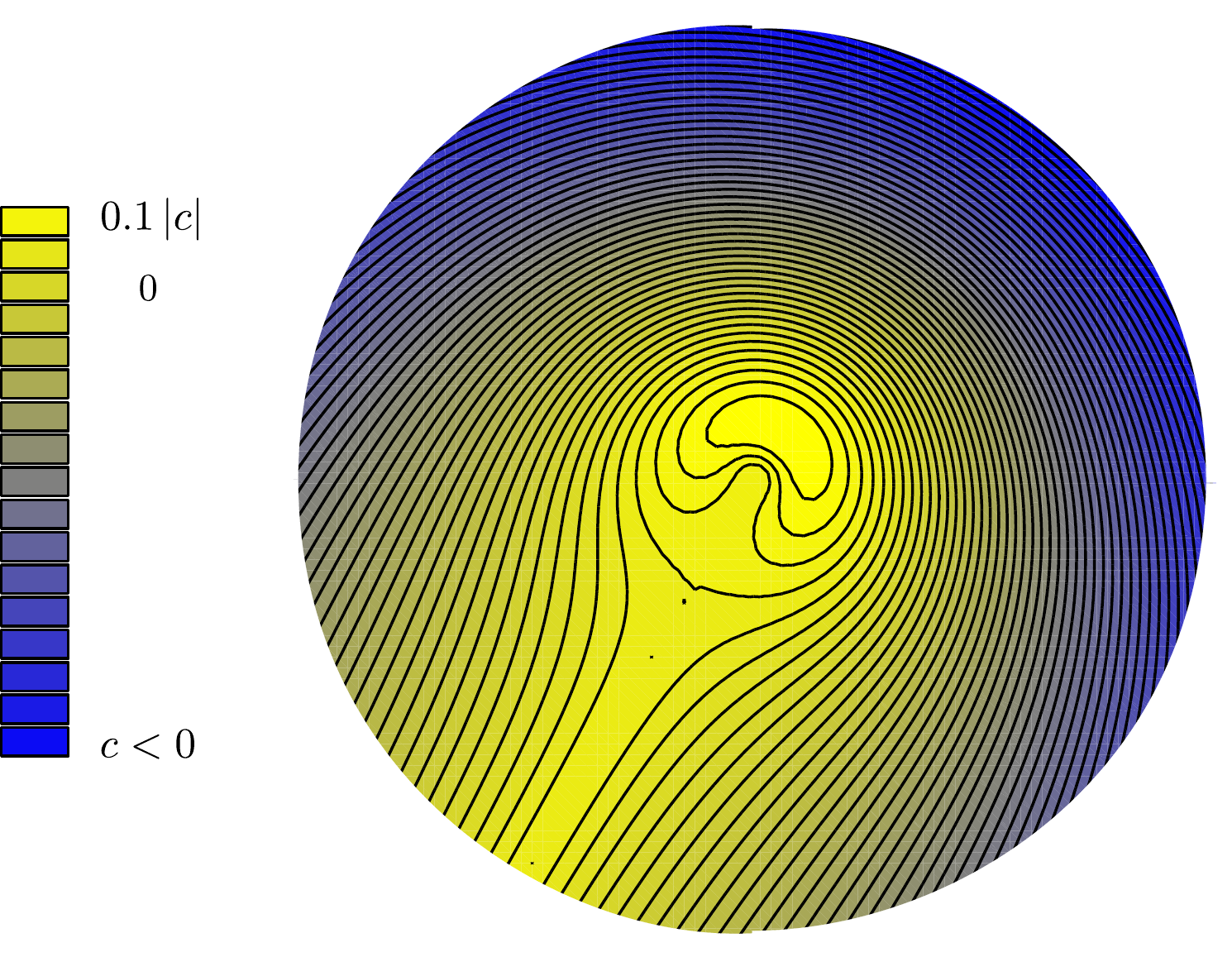}
        \caption{Contour plot of the buckling mode $\delta h_1$.}
        \label{fig:contour_3o2}
    \end{subfigure}
    \hfill
    \begin{subfigure}{0.54\textwidth}
        \centering
        \includegraphics[width=0.4\textwidth]{defect_plus_3o2-eps-converted-to.pdf}
        \caption{Field lines of $\m_0$ in equation \eqref{eq:n_0_q}.}
        \label{fig:lines_3o2}
        
         \vspace{0.5cm} 
        
        \includegraphics[width=\textwidth]{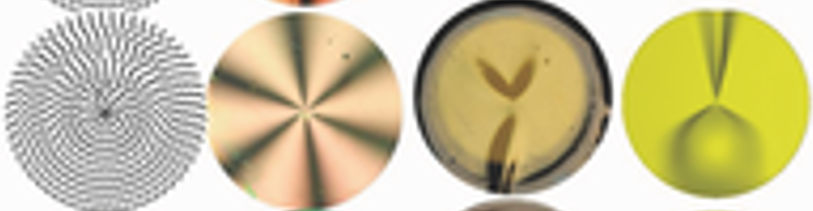}
        \caption{The $+3/2$ defect in LCN film \cite{McConney:2013:etal}.}
        \label{fig:delta_h_3o2}
     \end{subfigure}
    \caption{Illustration of the out-of plane buckling mode of deformation of LCN sheets with an imprinted director field $\m_0$ of topological charge $q=+\dfrac{3}{2}$ and $\vartheta_0=\dfrac{\pi}{3}$ in the reference configuration. For $\gamma=-1/2$, this occurs at the critical value $k\approx 1.43$.}
    \label{fig:main_3o2}
\end{figure}

\begin{figure}[htbp]
    \centering
    \begin{subfigure}{0.45\textwidth}
        \centering
        \vspace{1.05cm}
        \includegraphics[width=\textwidth]{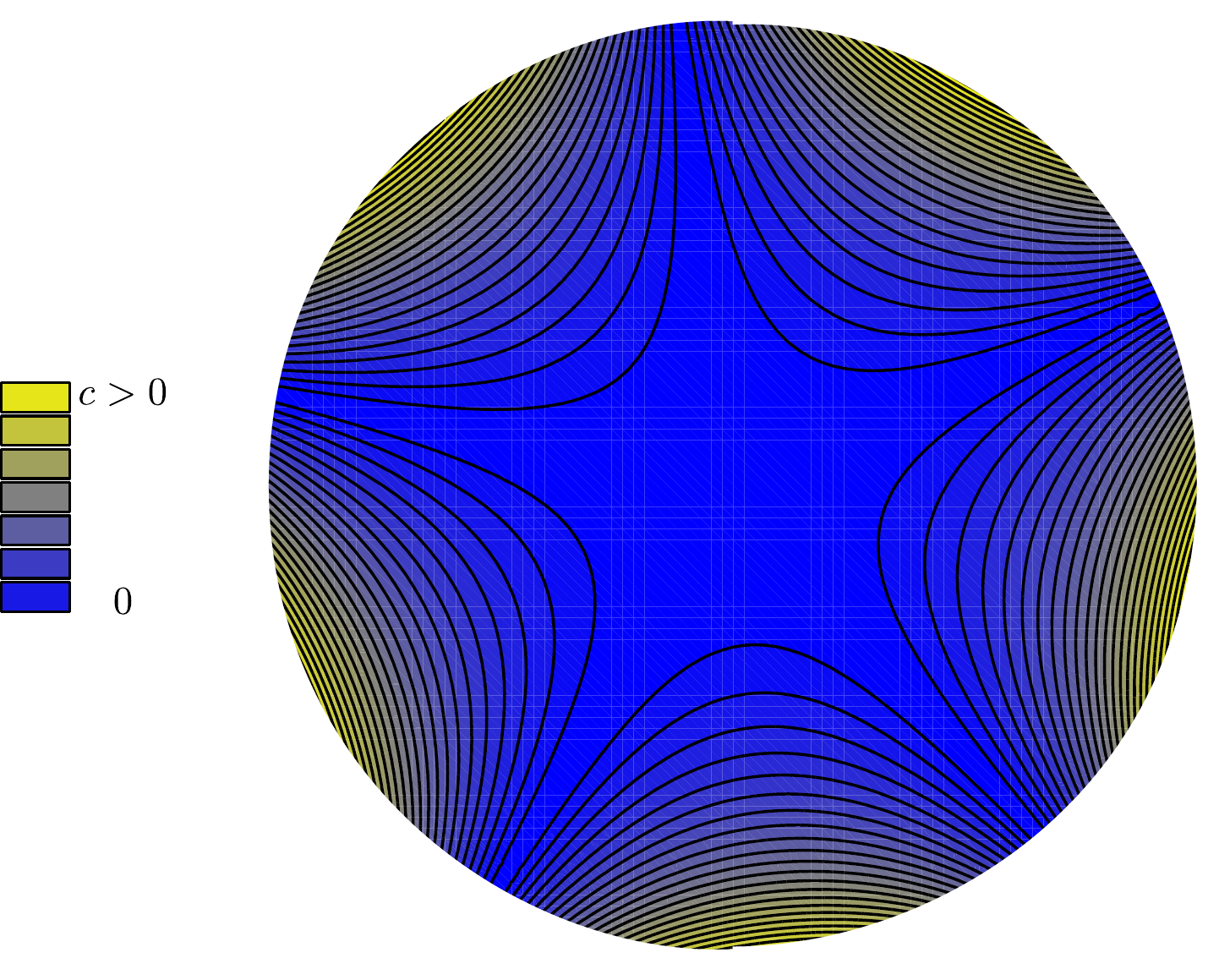}
        \caption{Contour plot of the buckling mode $\delta h_1$.}
        \label{fig:contour_minus_3o2}
    \end{subfigure}
    \hfill
    \begin{subfigure}{0.54\textwidth}
        \centering
        \includegraphics[width=0.4\textwidth]{defect_minus_3o2-eps-converted-to.pdf}
        \caption{Field lines of $\m_0$ in equation \eqref{eq:n_0_q}.}
        \label{fig:lines_minus3o2}
        
         \vspace{0.5cm} 
        
        \includegraphics[width=\textwidth]{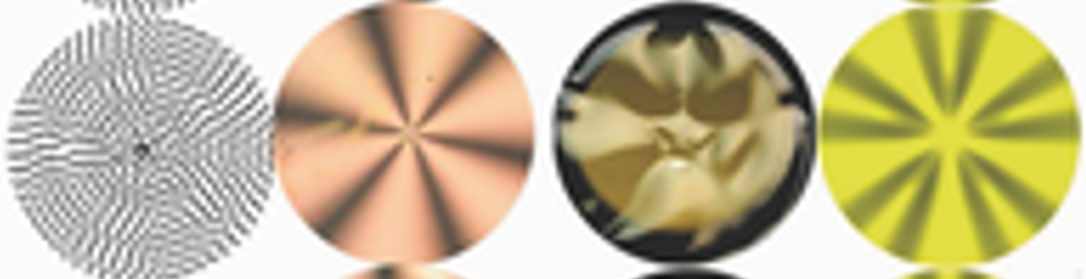}
        \caption{The $-3/2$ defect in LCN film \cite{McConney:2013:etal}.}
        \label{fig:delta_h_minus_3o2}
     \end{subfigure}
    \caption{Illustration of the out-of plane buckling mode of deformation of LCN sheets with an imprinted director field $\m_0$ of topological charge $q=-\dfrac{3}{2}$ and $\vartheta_0=\dfrac{\pi}{3}$ in the reference configuration. For $\gamma=-1/2$, this occurs at the critical value $k\approx 1.49$.}
    \label{fig:main_minus_3o2}
\end{figure}

\begin{figure}[htbp]
    \centering
    \begin{subfigure}{0.45\textwidth}
        \centering
        \vspace{0.8cm}
        \includegraphics[width=\textwidth]{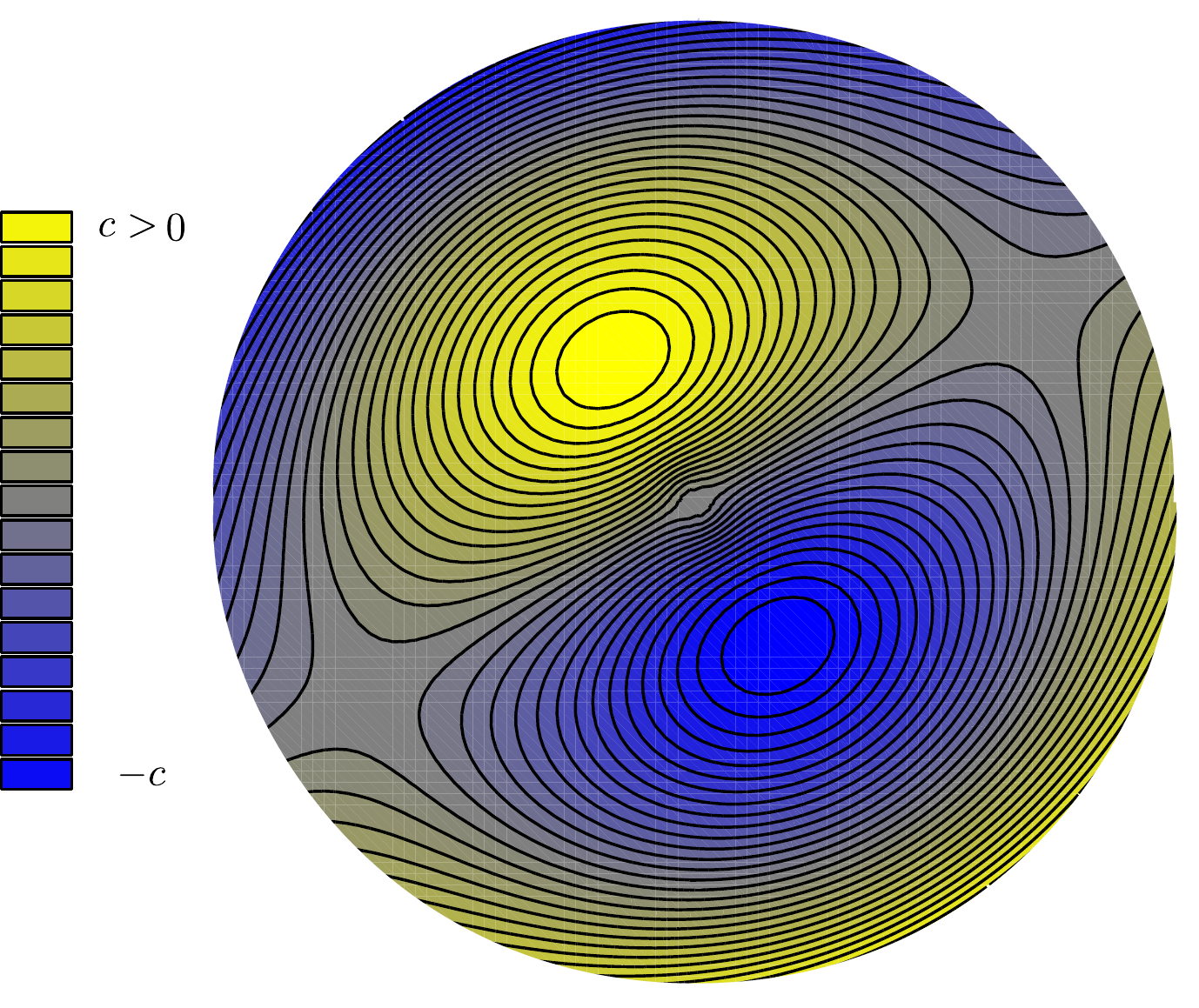}
        \caption{Contour plot of the buckling mode $\delta h_1$.}
        \label{fig:contour_2}
     \end{subfigure}
    \hfill
    \begin{subfigure}{0.54\textwidth}
        \centering
        \includegraphics[width=0.4\textwidth]{defect_plus_2-eps-converted-to.pdf}
        \caption{Field lines of $\m_0$ in equation \eqref{eq:n_0_q}.}
        \label{fig:lines_2}
        
         \vspace{0.5cm} 
        
        \includegraphics[width=\textwidth]{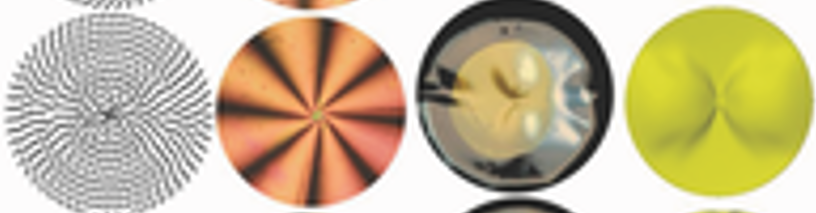}
        \caption{The $+2$ defect in LCN film \cite{McConney:2013:etal}.}
        \label{fig:delta_h_2}
    \end{subfigure}
    \caption{Illustration of the out-of plane buckling mode of deformation of LCN sheets with an imprinted director field $\m_0$ of topological charge $q=+2$ and $\vartheta_0=\dfrac{\pi}{3}$ in the reference configuration. For $\gamma=-1/2$, this occurs at the critical value $k\approx 0.73$.}
    \label{fig:main_2}
\end{figure}

\begin{figure}[htbp]
    \centering
    \begin{subfigure}{0.45\textwidth}
        \centering
        \vspace{0.8cm}
        \includegraphics[width=\textwidth]{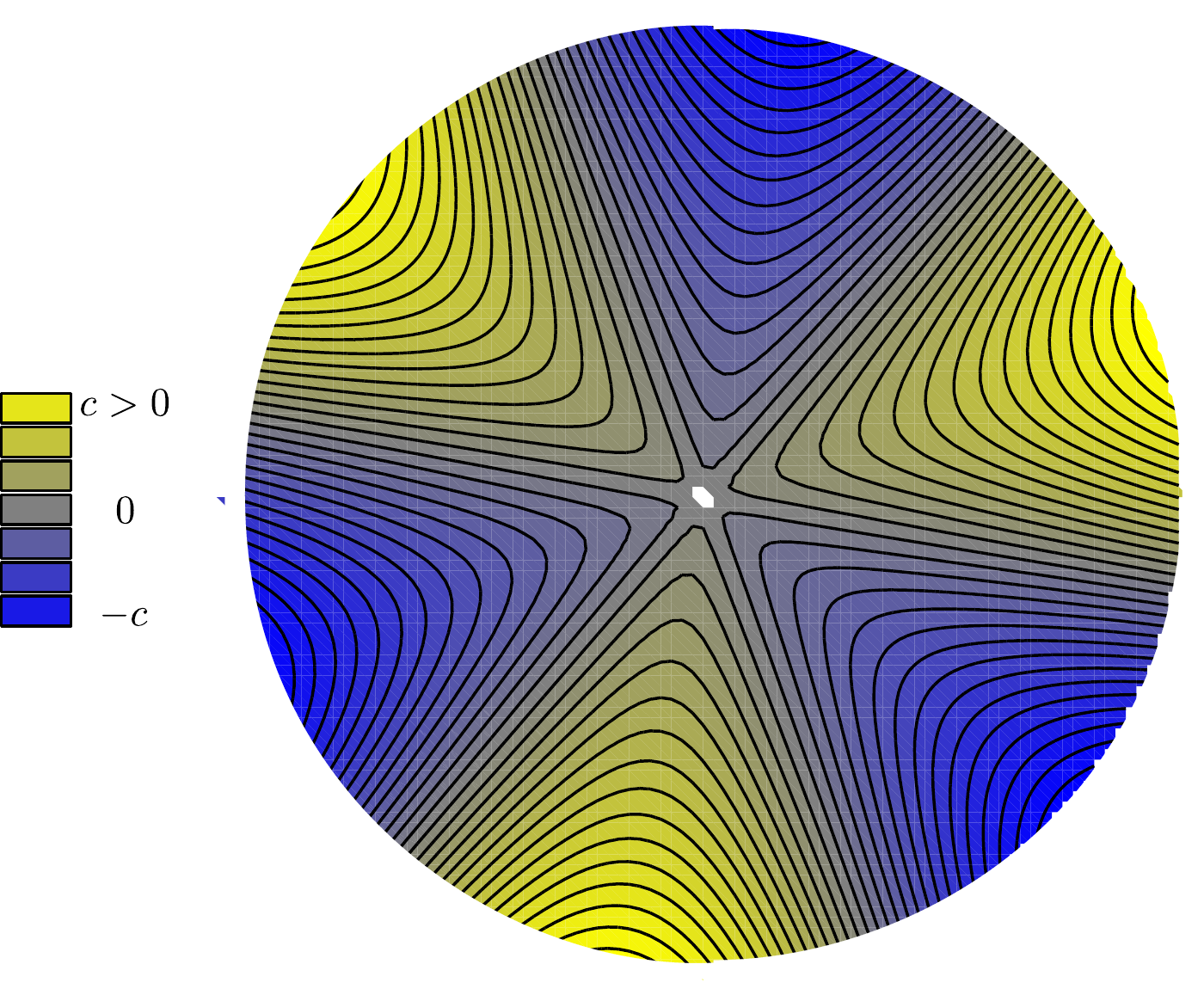}
        \caption{Contour plot of the buckling mode $\delta h_1$.}
        \label{fig:contour_minus_2}
     \end{subfigure}
    \hfill
    \begin{subfigure}{0.54\textwidth}
        \centering
        \includegraphics[width=0.4\textwidth]{defect_minus_2-eps-converted-to.pdf}
        \caption{Field lines of $\m_0$ in equation \eqref{eq:n_0_q}.}
        \label{fig:lines_minus_2}
        
         \vspace{0.5cm} 
        
        \includegraphics[width=\textwidth]{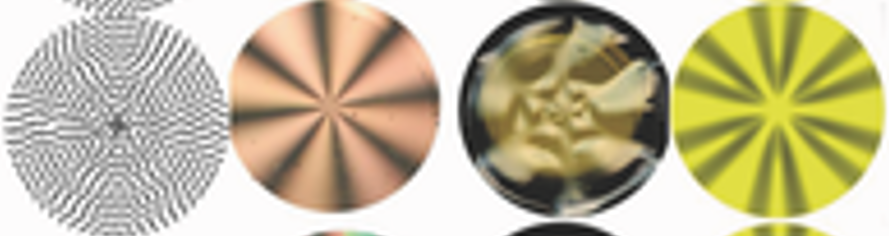}
        \caption{The $-2$ defect in LCN film \cite{McConney:2013:etal}.}
        \label{fig:delta_h_minus_2}
    \end{subfigure}
    \caption{Illustration of the out-of plane buckling mode of deformation of LCN sheets with an imprinted director field $\m_0$ of topological charge $q=-2$ and $\vartheta_0=\dfrac{\pi}{3}$ in the reference configuration. For $\gamma=-1/2$, this occurs at the critical value $k\approx 0.12$.}
    \label{fig:main_minus_2}
\end{figure}

\begin{figure}[htbp]
    \centering
    \begin{subfigure}{0.45\textwidth}
        \centering
        \vspace{0.98cm}
        \includegraphics[width=\textwidth]{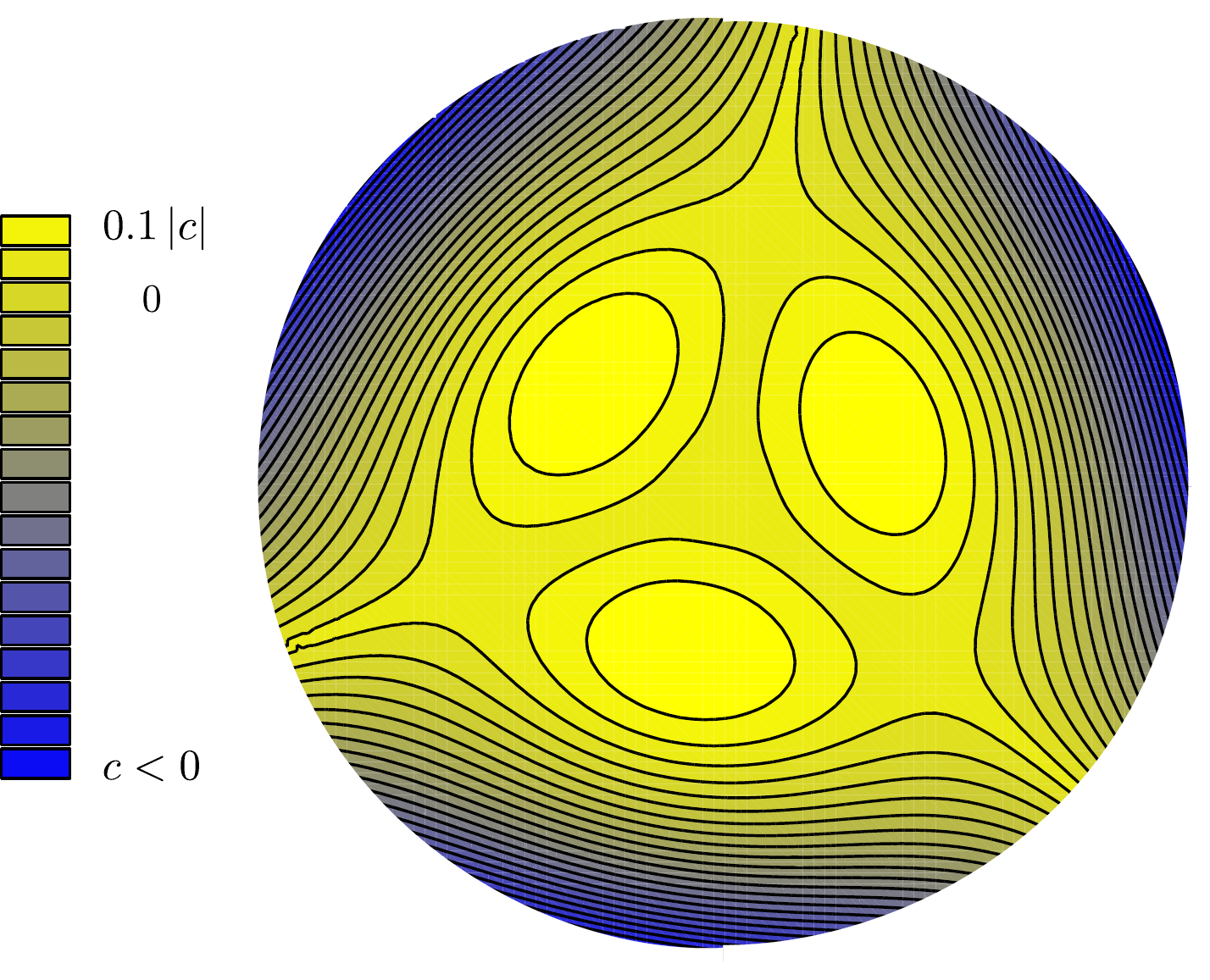}
        \caption{Contour plot of the buckling mode $\delta h_1$.}
        \label{fig:contour_5o2}
     \end{subfigure}
    \hfill
    \begin{subfigure}{0.54\textwidth}
        \centering
        \includegraphics[width=0.4\textwidth]{defect_5o2-eps-converted-to.pdf}
        \caption{Field lines of $\m_0$ in equation \eqref{eq:n_0_q}.}
        \label{fig:lines_5o2}
        
         \vspace{0.5cm} 
        
        \includegraphics[width=\textwidth]{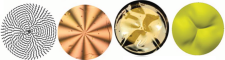}
        \caption{The $+5/2$ defect in LCN film \cite{McConney:2013:etal}.}
        \label{fig:delta_h_5o2}
    \end{subfigure}
    \caption{Illustration of the out-of plane buckling mode of deformation of LCN sheets with an imprinted director field $\m_0$ of topological charge $q=+\dfrac{5}{2}$ and $\vartheta_0=\dfrac{\pi}{3}$ in the reference configuration. For $\gamma=-1/2$, this occurs at the critical value $k\approx 1.39$.}
    \label{fig:main_5o2}
\end{figure}

\begin{figure}[htbp]
    \centering
    \begin{subfigure}{0.45\textwidth}
        \centering
        \vspace{0.8cm}
        \includegraphics[width=\textwidth]{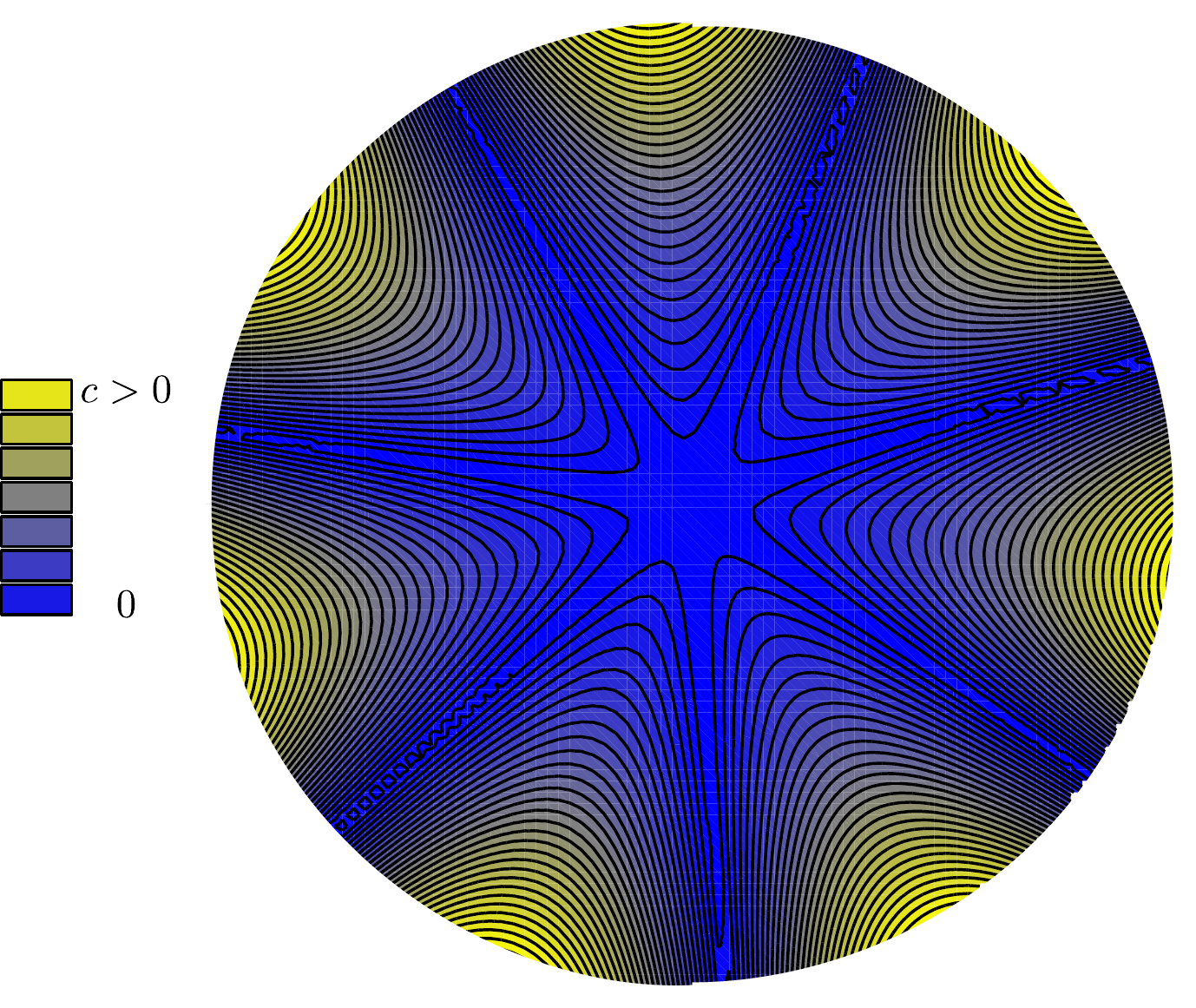}
        \caption{Contour plot of the buckling mode $\delta h_1$.}
        \label{fig:contour_minus_5o2}
     \end{subfigure}
    \hfill
    \begin{subfigure}{0.54\textwidth}
        \centering
        \includegraphics[width=0.4\textwidth]{defect_minus_5o2-eps-converted-to.pdf}
        \caption{Field lines of $\m_0$ in equation \eqref{eq:n_0_q}.}
        \label{fig:lines_minus_5o2}
        
         \vspace{0.5cm} 
        
        \includegraphics[width=\textwidth]{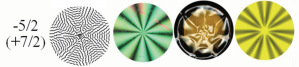}
        \caption{The $-5/2$ defect in LCN film \cite{McConney:2013:etal}.}
        \label{fig:delta_h_minus_5o2}
    \end{subfigure}
    \caption{Illustration of the out-of plane buckling mode of deformation of LCN sheets with an imprinted director field $\m_0$ of topological charge $q=-\dfrac{5}{2}$ and $\vartheta_0=\dfrac{\pi}{3}$ in the reference configuration. For $\gamma=-1/2$, this occurs at the critical value $k\approx 0.24$.}
    \label{fig:main_minus_5o2}
\end{figure}

In Figures~\ref{fig:contour_1o2}-\ref{fig:contour_minus_5o2}, we illustrate the contour plots of the buckling modes $\delta h_1=g(\vartheta)\bar f_1(r)$ for $q\in[-5/2,5/2]$, together with a comparison with the photomechanical response of azo-LCN films reported in \cite{McConney:2013:etal} and an illustration of the imprinted director field of the corresponding topological charge $q$ in the reference configuration. The critical value of $k$ for each $q$ is also indicated. As a multiplicative constant, we consider $c>0$ or $c<0$ to enhance the comparison with the experimental observation in \cite{McConney:2013:etal}; since we are considering out-of-plane deformation of an initially flat disk, upward and downward displacements are energetically equivalent.

The deformation of the $+1/2$ defect is characterized as a valley (on which $\delta h_1$ is zero) extending outward from the defect center toward the film's edge along the direction determined by the defect orientation $\bm p$. In the remaining region, the flat configuration undergoes an out-of-plane deformation, with a peak positioned opposite to the valley. The
deformation of the $+1$ defect can be described as a round cone, very similar to the predicted
deformation \cite{modes2010:disclination}. The $+3/2$ defects autonomously deforms into a tear-shaped dimple morphology and a valley region (along the defect orientation $\bm p$),
both radiating from opposite side of the center of the defect. The $+2$ defect, with 2-fold
symmetry, deforms into two tear shaped dimples radiating from opposite sides of the center of
the defect, while the $+5/2$ defect, with 3-fold symmetry has 3 tear-shaped
dimples radiating from the center. The self-molded shape of the $-1/2$ defect can be described as
three valleys radiating from the center at $120^\circ$ offset with respect to each other. The
photo-induced deformation of the $-1$, $-3/2$, $-2$, and $-5/2$ defects are made up of four, five, six, and seven displacements between valleys and peaks pointing toward the center, respectively.

\section{Conclusion}\label{sec:conclusion}

Different from other polymer networks, topological defects in liquid crystal networks (LCNs) give rise to richer shape-morphing modes and mechanisms. They are also more diverse and their evolution much richer than in fluid LCs.

We present here a clear mathematical description of textures caused by topological defects in LCNs by focusing on a class of equilibrium configurations represented by a thin LCN sheet containing various central defects and subject to an external stimulus. In the experimental study \cite{McConney:2013:etal}, similar LCN samples, prepared with an imprinted central defect of topological charge $q$, deformed by adopting a complex topography specific to that defect. 

In our model, macroscopic deformations arise from the material's response to two distinct mechanisms: (i) the mechanical stress induced by the distortion of the director field in the reference configuration and (ii) variations in the nematic order parameter. The first activation mechanism is captured by the nematic contribution to the free energy, which does not retain memory of the initial degree of orientation but quantifies the energetic cost associated with the distortion of the director field in the deformed configuration. The second contribution to the free energy is based on hyperelastic models for nematic polymer networks \cite{Mihai:2020b:MG,sonnet2022:photoresponsive} and accounts for how variations in the degree of orientation drive the system out of equilibrium, leading to shape changes. We derive general equilibrium and stability conditions by analyzing the first and second variations of the energy functional.

Our results suggest that two key parameters govern the buckling instability of the thin sheet: a parameter $\gamma>-2$, which represents the tendency of monomers in polymer chains to distribute anisotropically in response to an external stimulus (such as heat or light), and a parameter $k>0$, interpreted as a reduced elastic constant that balances nematic elasticity, material stiffness, and variations in the nematic order $s$. For a given admissible $\gamma$, we determined the critical value of $k$ above which the flat configuration ceases to be stable against a restricted set of perturbations. This provides an upper bound for the value of $k$ at which the system becomes unstable. Furthermore, we characterized the corresponding buckling modes, which share the periodicity of the defect imprinted in the LCN. Specifically, for a defect of topological charge $q$, we identify $2|q-1|$ periodic regions where the direction of the director repeats itself, and these regions of periodicity characterize also the deformation. Our results are in good qualitative agreement with the available experimental examples.

These findings open several avenues for further exploration. A follow-up question concerns the role of boundary conditions in stabilizing the flat ground state. Specifically, one can ask whether there is a class of anchoring conditions capable of preventing out-of-plane deformation of LCN sheets for all $k>0$ and $\gamma>-2$. The study of interactions between neighboring defects \cite{Tang:2017:TS} and their influence on shape deformations presents another intriguing direction for future research.

Our framework could also be extended to different geometries of the initial configuration, or to incorporate the evolution of orientational order, hydrodynamic coupling, active defect dynamics, and shape-changing surfaces.

Furthermore, our model can form the basis for a mathematical understanding of how developing organisms achieve biological organization by growing persistent protrusions induced by topological defects. In the context of morphogenesis, our setup can be interpreted as an initially isotropic phase where cells adopt a specific degree of orientation dictated by $s$, forming defects at locations where $s$ vanishes. This transition represents an active impulse that remains to be understood.

\subsection*{Acknowledgments} The work of S.P. and G.G.G. is partially supported by the European Union–Next Generation EU under the National Recovery and Resilience Plan (NRRP), Mission 4 Component 2 Investment 1.1–Call PRIN 2022 of Italian Ministry of University and Research. Project No. 202249PF73
``Mathematical models for viscoelastic biological matter''. L.A.M. and S.P. also thank the Isaac Newton Institute for Mathematical Sciences (INI), Cambridge, for their support and hospitality during the research programme ``Uncertainty Quantification and Stochastic Modelling of Materials'' (USM), July-September 2023, when discussions about  these were initiated. The USM programme was supported by EPSRC Grant Number EP/R014604/1.

\appendix

\section{First and second variations: Detailed calculations}\label{sec:first_second_var_app}
\setcounter{equation}{0}
\renewcommand{\theequation}{A.\arabic{equation}}

To compute the first variation $\delta_1\mathcal{F}[\bm\varphi]$ of our rescaled free-energy functional $\mathcal{F}[\bm\varphi]$ in \eqref{eq:free_rescaled}, we introduce a test field $\bG_1=\delta_1\bF$ that represents the variation of the deformation gradient $\bF$ in a way compatible with the incompressibility constraint $\det\bF=1$. This is possible by ensuring that the following equality holds,
\begin{equation}
\label{eq:incompressibility_delta}
\delta_1\left(\det\bF\right)=0.
\end{equation}
Here, $\delta_1$ denotes the first variation applied to the functional $\mathcal{F}$, represented by $\bG_1$. The second variation $\delta_2\delta_1\mathcal{F}$ is obtained by further perturbing $\delta_1\mathcal{F}$ through another admissible variation $\bG_2$ of $\bF$.
Using the definitions of determinant and adjugate $\bF^\ast$ of the tensor $\bF$, and considering an orthonormal basis $(\e_1,\e_2,\e_3)$ of $\mathcal{V}$, \eqref{eq:incompressibility_delta} can be rewritten as
\begin{equation}
\label{eq:incompressibility_delta2}
\delta_1(\bF\e_1\cdot\bF\e_2\times\bF\e_3)=\tr({\bF^{\ast}}^{\mathrm{T}}\bG_1)=\det\bF\tr\left(\bF^{-1}\bG_1\right)=0,
\end{equation}
and so all the admissible perturbations $\bG_1$ of $\bF$ satisfies \eqref{eq:perturb_admissible}. The same holds for the admissible $\bG_2$. Since $\n$ is delivered by the deformation through \eqref{eq:n_deformed}, we can also introduce the test field $\vv_1$ that represents the variation of $\n$ through $\bG_1$,
\begin{equation}
\label{eq:vv_1_def}
\vv_1=\delta_1\n=\frac{1}{|\bF\n_0|}\bP(\n)\left(\bG_1\n_0\right),
\end{equation}
where $\bP(\n)=\bm{\mathrm{I}}-\n\otimes\n$ is the projection onto the plane orthogonal to $\n$. We note that $\vv_1$ is orthogonal to $\n$, which is consistent with the constraint of unimodularity for $\n$, $\delta_1(\n\cdot\n)=0$. 
Since
\begin{subequations}
\label{eq:helping_first_variation}
\begin{align}
\delta_1\left|\nabla_{\bm X}\n\bm{\mathrm{F}}^{-1}\right|^2=2\left[(\nabla\vv_1)\bF^{-1}+\nabla\n \,\delta_1(\bF^{-1})\right]\cdot\nabla\n\bF^{-1}=&2\left[(\nabla\vv_1)\bF^{-1}-\nabla\n\bF^{-1}\bG_1\bF^{-1}\right]\cdot\nabla\n\bF^{-1}\label{eq:helping_first_variation1}\\
\delta_1\tr\left(\bm{\mathrm{F}}^{\mathrm{T}}\bm{\mathrm{F}}\right)=&2\bG_1\cdot\bF
\end{align}
\end{subequations}
In \eqref{eq:helping_first_variation1} use has been made of the fact that since $\bF^{-1}\bF=\bm{\mathrm{I}}$, then $\delta_1(\bF^{-1})\bF+\bF^{-1}\bG_1=\bm 0$, and so 
\begin{equation}
\label{eq:delta_F_minus_1}
\delta_1(\bF^{-1})=-\bF^{-1}\bG_1\bF^{-1}.
\end{equation}
Moreover, given the definition of $\bL$ as a function of $\n$ in \eqref{eq:step_tensors}, it follows that the constraint $\det\bL=1$ is preserved under the perturbation. Indeed, by introducing the test field $\bm{\mathrm{M}}$, which represents the first variation of $\bL$ through $\bG_1$ and is defined as
\begin{equation}
\label{eq:delta_1_L}
\bm{\mathrm{M}}_1=\delta_1\bL=a^{-1/3}(a-1)(\vv_1\otimes\n+\n\otimes\vv_1),
\end{equation}
with $\vv_1$ defined in \eqref{eq:vv_1_def}, we compute
\begin{equation}
\label{eq:detL_perturb}
\delta_1(\det\bL)=\tr\left(\bL^{-1}\bm{\mathrm{M}}_1\right)=a^{-1/3}(a-1)\tr\left(a^{-1}\n\otimes\vv_1+\vv_1\otimes\n\right),
\end{equation}
which vanishes due to the orthogonality of $\n$ and $\vv$. Moreover, since $\bL^{-1}\bL=\bm{\mathrm{I}}$, then 
\begin{equation}
\label{eq:delta_L_minus_1}
\delta_1\bL^{-1}=-\bL^{-1}\bm{\mathrm{M}}_1\bL^{-1}=-a^{1/3}\left(a^{-1}-1\right)(\n\otimes\vv_1+\vv_1\otimes\n).
\end{equation}
The first variation of the dimensionless energy functional $\mathcal{F}$ at the field $\bm\varphi$ is then a linear functional of $\bG_1$ and results in \eqref{eq:stability_condition}.

By further perturbing $\delta_1\mathcal{F}$, we compute the second variation of the functional $\mathcal{F}$. Now $\bG_2=\delta_2\bF$ represents another admissible perturbation of $\bF$, such that $\bG_2\cdot\bF^{-\mathrm{T}}=0$, and we define $\vv_2$ and $\bm{\mathrm{M}}_2$ the test fields that represents the variation of $\n$ and $\bL$, respectively, through $\bG_2$,
\begin{subequations}
\begin{align}
\vv_2&=\delta_2\n=\frac{1}{|\bF\n_0|}\bP(\n)\left(\bG_2\n_0\right),\label{eq:vv_2_def}\\
\bm{\mathrm{M}}_2&=\delta_2\bL=a^{-1/3}(a-1)(\vv_2\otimes\n+\n\otimes\vv_2).\label{eq:delta_2_L}
\end{align}
\end{subequations}
Moreover, since also $\vv_1$ in \eqref{eq:vv_1_def} depends on $\bF$, we denote by $\bm\xi=\delta_2\vv_1$ its variation through $\bG_2$, which results in 
\begin{equation}
\label{eq:xi_def}
\bm\xi=\delta_2\vv_1=-\frac{1}{|\bF\n_0|}\left[(\vv_1\otimes\n)\bG_2\n_0+(\vv_2\otimes\n+\n\otimes\vv_2)\bG_1\n_0\right]
\end{equation}
Since, as before, $\delta_2(\bF^{-1})=-\bF^{-1}\bG_2\bF^{-1}$, we can express the second variation $\delta_2\delta_1\mathcal{F}$ at the field $\bm\varphi$ as in \eqref{eq:stability_condition}, which is a bilinear form in $\bG_1$ and $\bG_2$.

\section{Out-of-plane deformations: Detailed calculations}\label{sec:out_of_plane_app}
\setcounter{equation}{0}
\renewcommand{\theequation}{B.\arabic{equation}}

We apply \eqref{eq:equilibrium_condition} to $\bm\varphi_0$ in \eqref{eq:phi_flat}, with $\n_0=\m_0$ in \eqref{eq:n_0_q}, describing a central defect with topological charge $q$, and $s$ given by \eqref{eq:s_out_of_plane}. By rescaling all lengths to the radius $R$ of the initial cylinder $\body_0$ in \eqref{eq:flat_LCN}, instead of to $\diam$, \eqref{eq:equilibrium_condition} reduces to
\begin{align}
\label{eq:equilibrium_condition_flat}
\delta_1\mathcal{F}[\bm\varphi_0](\bG_1)&=\frac{H}{R}\int_\varepsilon^1\int_0^{2\pi}\left[2\kappa \bar{s}^2\left(-\nabla\vv_1|_{\bm\varphi_0}+\nabla\m_0\bG_1\right)\cdot\nabla\m_0+\tr\left(\bG_1^{\mathrm{T}}\bL^{-1}\bL_0\right)|_{\bm\varphi_0}\right.\nonumber\\
&\left.+\tr\left(\bL^{-1}\bG\bL_0\right)|_{\bm\varphi_0}+\bar{a}^{1/3}\left(\bar{a}^{-1}-1\right)\tr\left((\m_0\otimes\vv_1|_{\bm\varphi_0}+\vv|_{\bm\varphi_0}\otimes\m_0)\bL_0\right)\right]r\dd r\dd\vartheta,
\end{align}
where $\nabla$ represents the gradient in polar coordinates, while the perturbation $\vv_1|_{\bm\varphi_0}$ of $\m_0$ through $\bG_1$ results by \eqref{eq:vv_1_def} to be
\begin{equation}
\label{eq:vv_1_flat}
\vv_1|_{\bm\varphi_0}=v_1\e_z, \quad \hbox{ where } \, v_1=\nabla\delta h_1\cdot\m_0=\delta h_{1,r}\cos w+\frac{1}{r}\delta h_{1,\vartheta}\sin w.
\end{equation}
Moreover,
\begin{equation}
\label{eq:nabla_n_0_q}
\nabla\m_0=\frac{q}{r}\m_{0\perp}\otimes\e_\vartheta, \quad \hbox{ where } \, \m_{0\perp}=-\sin w\e_r+\cos w\e_\vartheta,
\end{equation}
with $w=w(\vartheta)$ defined in \eqref{eq:n_0_q}, and $\bar a_0$ and $\bar a$ are functions of $\bar s_0$ and $\bar s$, respectively, and thus remain constant. Since 
\begin{equation}
\label{eq:auxiliary_equilibrium_flat}
\tr\left(\bG_1^{\mathrm{T}}\bL^{-1}\bL_0\right)|_{\bm\varphi_0}=\tr\left(\bL^{-1}\bG\bL_0\right)|_{\bm\varphi_0}=a^{1/3}a_0^{-1/3}\nabla\delta_1h\otimes\e_z,
\end{equation}
and since $\m_0$, $\m_{0\perp}$ and $\nabla\delta h_1$ are orthogonal to $\e_z$, it follows from \eqref{eq:vv_1_flat}, \eqref{eq:nabla_n_0_q} and \eqref{eq:variation_out_of_plane} that $\delta_1\mathcal{F}[\varphi_0](\bG_1)$ in \eqref{eq:equilibrium_condition_flat} vanishes for all admissible test fields $\bG_1$. Therefore, $\varphi_0$ is an equilibrium configuration, and we will study its stability.

The second variation of $\mathcal{F}$ in \eqref{eq:stability_condition} at the equilibrium configuration $\bm\varphi_0$ in \eqref{eq:phi_flat} is given by
\begin{align}
\label{eq:stability_condition_flat_app}
&\delta_1\delta_2\mathcal{F}[\bm{\varphi}_0](\bG_1,\bG_2)=\frac{H}{R}\int_\varepsilon^1\int_0^{2\pi}\left\{2\kappa \bar{s}^2\left[\left(\nabla\bm\xi|_{\bm\varphi_0}-\bm{\mathrm{A}}[\vv_1|_{\bm\varphi_0},\bm{\mathrm{G}}_1]\bm{\mathrm{G}}_2-\bm{\mathrm{A}}[\vv_2|_{\bm\varphi_0},\bm{\mathrm{G}}_2]\bm{\mathrm{G}_1}\right)\cdot(\nabla\m_0)\right.\right.\nonumber\\
&\left.\left.+\bm{\mathrm{A}}[\vv_1|_{\bm\varphi_0},\bm{\mathrm{G}}_1]\cdot\bm{\mathrm{A}}[\vv_2|_{\bm\varphi_0},\bm{\mathrm{G}}_2]\right]+\tr\left(\bG_1^{\mathrm{T}}\bL^{-1}\bG_2\bL_0\right)|_{\bm\varphi_0}+\tr\left(\bG_2^{\mathrm{T}}\bL^{-1}\bG_1\bL_0|_{\bm\varphi_0}\right)\right.\nonumber\\
&\left.+\left(\bar{a}^{-1/3}-\bar{a}^{1/3}\right)\left[\tr\left(\bG_1^{\mathrm{T}}(\m_0\otimes\vv_2|_{\bm\varphi_0}+\vv_2|_{\bm\varphi_0}\otimes\m_0)\bL_0|_{\bm\varphi_0}\right)\right.\right.\nonumber\\
&\left.\left.+\tr\left(\bG_2^{\mathrm{T}}(\m_0\otimes\vv_1|_{\bm\varphi_0}+\vv_1|_{\bm\varphi_0}\otimes\m_0)\bL_0|_{\bm\varphi_0}\right)\right.\right.\nonumber\\
&\left.\left.+\tr\left((\m_0\otimes\vv_2|_{\bm\varphi_0}+\vv_2|_{\bm\varphi_0}\otimes\m_0)\bG_1\bL_0|_{\bm\varphi_0}\right)+\tr\left((\m_0\otimes\vv_1|_{\bm\varphi_0}+\vv_1|_{\bm\varphi_0}\otimes\m_0)\bG_2\bL_0|_{\bm\varphi_0}\right)\right.\right.\nonumber\\
&\left.\left.+\tr\left((\vv_2|_{\bm\varphi_0}\otimes\vv_1|_{\bm\varphi_0}+\vv_1|_{\bm\varphi_0}\otimes\vv_2|_{\bm\varphi_0})\bL_0|_{\bm\varphi_0}\right)+\tr\left((\m_0\otimes\xi|_{\bm\varphi_0}+\xi|_{\bm\varphi_0}\otimes\m_0)\bL_0|_{\bm\varphi_0}\right)\right]\right\}r\dd r\dd\vartheta,
\end{align}
where $\vv_1|_{\bm\varphi_0}$ is defined in \eqref{eq:vv_1_flat} and
\begin{subequations}
\label{eq:identities_second_var}
\begin{align}
\vv_2|_{\bm\varphi_0}&=v_2\e_z, \quad \quad \hbox{ where } \, v_2=\nabla\delta h_2\cdot\m_0=\delta h_{2,r}\cos w+\frac{1}{r}\delta h_{2,\vartheta}\sin w,\\
\bm\xi|_{\bm\varphi_0}&=-\left[(\vv_1|_{\bm\varphi_0}\otimes\m_0)\bG_2\m_0+(\vv_2|_{\bm\varphi_0}\otimes\m_0+\n\otimes\vv_2|_{\bm\varphi_0})\bG_1\m_0\right]=-v_1v_2\m_0.
\end{align}
\end{subequations}
Making use of \eqref{eq:vv_1_flat}, \eqref{eq:identities_second_var} and \eqref{eq:variation_out_of_plane}, we obtain that
\begin{subequations}
\label{eq:identities_second_var1}
\begin{align}
\nabla\bm\xi|_{\bm\varphi_0}\cdot\nabla\m_0=&-\frac{q^2}{r^2}v_1v_2,\\
\bm{\mathrm{A}}[\vv_1|_{\bm\varphi_0},\bm{\mathrm{G}}_1]\bm{\mathrm{G}}_2\cdot\nabla\m_0=\bm{\mathrm{A}}[\vv_2|_{\bm\varphi_0},\bm{\mathrm{G}}_2]\bm{\mathrm{G}_1}\cdot\nabla\m_0=& \,0,\\
\bm{\mathrm{A}}[\vv_1|_{\bm\varphi_0},\bm{\mathrm{G}}_1]\cdot\bm{\mathrm{A}}[\vv_2|_{\bm\varphi_0},\bm{\mathrm{G}}_2]=&\nabla v_1\cdot\nabla v_2,
\end{align}
\end{subequations}
and $\delta_2\delta_1\mathcal{F}[\bm\varphi_0]$ in \eqref{eq:stability_condition_flat_app} reduces to \eqref{eq:second_variation_out_of_plane} in the main text.
We recall that $\nabla=\partial_r+\frac{1}{r}\partial_\vartheta+\partial_z$ is the gradient in polar coordinates.

\section{Determination of buckling modes}\label{sec:solution_app}

By substituting \eqref{eq:separation_of_variables} and \eqref{eq:ansatz_g_i} in \eqref{eq:second_variation_out_of_plane}, the integrals over $r$ and $\vartheta$ decouple. Upon integrating by parts with respect to $\vartheta$, and since $g'(2\pi)=\pm g'(0)$ according to $q$ and $\vartheta_0$, we obtain
\begin{align}
\label{eq:second_variation_out_of_plane_separation}\delta_2\delta_1\mathcal{F}[\bm\varphi_0](f_1,f_2)&=\frac{H}{R}\int_\varepsilon^1\left\{k(A-A_w)f_1''f_2''+\frac{ k}{2}(q-1)(A-2A_w)\left[\frac{1}{r}\left(f_1f_2''+f_1''f_2\right)-f_1'f_2''-f_1''f_2'\right]\right.\nonumber\\
&\left.+\frac{1}{r}f_1'f_2'\left[ k\left((A-A_w)(-2q+1)+B\right)+(A+\gamma(A-A_w))r^2\right]\right.\nonumber\\
&\left.+\frac{1}{r^3}f_1f_2\left[ k\left(-2B_w(q-1)+C_w\right)+(B+\gamma B_w)r^2\right]\right.\nonumber\\
&\left.+\frac{1}{2r^2}(f_1f_2'+f_1'f_2)\left[ k\left((A-2A_w)(q-1)(2q-1)+2B_w(q-2)+B(q-1)\right)\right.\right.\nonumber\\
&\left.\left.\qquad-\gamma(A-2A_w)(q-1)r^2\right]\right\}\dd r,
\end{align}
where a prime denotes differentiation, $k$ and $\gamma$ are defined in \eqref{eq:k_gamma_elastomeric}, and
\begin{equation}
\label{eq:A_Aw_B_Bw_Cw}
\begin{split}
&A=\int_0^{2\pi}g^2\dd\vartheta, \qquad A_w=\int_0^{2\pi}g^2\sin^2w\dd\vartheta, \\
&B=\int_0^{2\pi}g'^2\dd\vartheta, \qquad B_w=\int_0^{2\pi}g'^2\sin^2w\dd\vartheta, \\
&C_w=\int_0^{2\pi}g''^2\sin^2w\dd\vartheta,
\end{split}
\end{equation}
with $g$ defined in \eqref{eq:ansatz_g_i}. 

Integrating by parts with respect to $r$ in \eqref{eq:second_variation_out_of_plane_separation} yields
\begin{align}
\label{eq:weak_form_EL}
&\delta_2\delta_1\mathcal{F}[\bm\varphi_0](f_1,f_2)=\frac{H}{R}\int_\varepsilon^1\left\{k r(A-A_w)f_1^{iv}+2 k (A-A_w)f_1'''\right.\nonumber\\
&\left.+\frac{f_1''}{r}\left[ k \left((3q-2)A-(4q-3)A_w-B\right)-(A+\gamma(A-A_w))r^2\right]\right.\nonumber\\
&\left.+\frac{f_1'}{r^2}\left[- k \left((3q-2)A-(4q-3)A_w-B\right)-(A+\gamma(A-A_w))r^2\right]\right.\nonumber\\
&\left.+\frac{f_1}{r^3}\left[ k \left(2q(q-1)(A-2A_w)+B(q-1)+C_w-2B_w\right)+(B+\gamma B_w)r^2\right]\right\}f_2\dd r\nonumber\\
&+\left.\left[- r(A-A_w)f_1'''+\frac{f_1'}{2r}\left[ k \left(A(-5q+4)+A_w(6q-4)+2B\right)+2(A+\gamma(A-A_w))r^2\right]\right.\right.\nonumber\\
&\left.\left.+\frac{f_1}{2r^2}\left[ k ((A-2A_w)(q-1)(2q-1)+2B_w(q-2)+B(q-1))+\gamma(2A_w-A)(q-1)r^2\right]\right]\right|_{r=1}f_2(1)\nonumber\\
&\left.+ k \left[r(A-A_w)f_1''+\frac{(q-1)}{2}(A-2A_w)\left(\frac{1}{r}f_1-f'_1\right)\right]\right|_{r=1}f_2'(1)\nonumber\\
&- k \left.\left[r(A-A_w)f_1''+\frac{(q-1)}{2}(A-2A_w)\left(\frac{1}{r}f_1-f'_1\right)\right]\right|_{r=\varepsilon}f_2'(\varepsilon).
\end{align}
The equations obeyed by a generic variation $f_1$ satisfying \eqref{eq:conditions_fg} readily follows from \eqref{eq:weak_form_EL}. The integrals and the three quantities evaluated at $r=1$ or $r=\varepsilon$ are independent of each other. To make the second variation equal to $0$, these quantities must vanish for all variations $f_2$. Therefore, due to the arbitrariness of $f_2$, and the values $f_2(1)$, $f_2'(1)$ and $f_2'(0)$, $f_1$ must then satisfy equations \eqref{eq:conditions_f1_null_second_variation}.

A generic solution $f_1$ of equation \eqref{eq:bulk_eq} in the bulk for $q=\dfrac{2m+1}{2}$, $m\in\mathbb{Z}$ and $g(\vartheta)=|\sin w|$ as in \eqref{eq:ansatz_g_i} is given by
\begin{equation}
\label{eq:f_1el_sin}
    f_1(r)=f_{11}(r)+f_{12}(r)+f_{13}(r)+f_{14}(r),
\end{equation}
where
\begin{align}
&f_{11}(r)=\mathit{c_1}r^{1-\frac{\alpha_1}{2}}\mathrm{hypergeom}\! \left(\left[\frac{1}{4}-\frac{\alpha_1}{4},\frac{3}{4}-\frac{\alpha_1}{4}\right],\left[1-\frac{\alpha_1}{2},1+\frac{\alpha_2}{4}-\frac{\alpha_1}{4},1-\frac{\alpha_2}{4}-\frac{\alpha_1}{4}\right],\frac{r^{2}\left(4+\gamma \right)}{4 k}\right)\nonumber,
\end{align}
\begin{align}
&f_{12}(r)=\mathit{c_2}r^{1+\frac{\alpha_1}{2}}\mathrm{hypergeom}\! \left(\left[\frac{1}{4}+\frac{\alpha_1}{4},\frac{3}{4}+\frac{\alpha_1}{4}\right],\left[1+\frac{\alpha_1}{2},1+\frac{\alpha_2}{4}+\frac{\alpha_1}{4},1-\frac{\alpha_2}{4}+\frac{\alpha_1}{4}\right],\frac{r^{2}\cdot \left(4+\gamma \right)}{4k}\right)\nonumber,
\end{align}
\begin{align}
&f_{13}(r)=\mathit{c_3}r^{1-\frac{\alpha_2}{2}} \mathrm{hypergeom}\! \left(\left[\frac{1}{4}-\frac{\alpha_2}{4},\frac{3}{4}-\frac{\alpha_2}{4}\right],\left[1-\frac{\alpha_2}{2},1-\frac{\alpha_2}{4}+\frac{\alpha_1}{4},1-\frac{\alpha_2}{4}-\frac{\alpha_1}{4}\right],\frac{r^{2}\left(4+\eta \right)}{4 k}\right)\nonumber,
\end{align}
\begin{align}
&f_{14}(r)=\mathit{c_4}r^{1+\frac{\alpha_2}{2}} \mathrm{hypergeom}\! \left(\left[\frac{1}{4}+\frac{\alpha_2}{4},\frac{3}{4}+\frac{\alpha_2}{4}\right],\left[1+\frac{\alpha_2}{2},1+\frac{\alpha_2}{4}+\frac{\alpha_1}{4},1+\frac{\alpha_2}{4}-\frac{\alpha_1}{4}\right],\frac{r^{2}\cdot \left(4+\gamma \right)}{4k}\right),\nonumber
\end{align}
with
\begin{equation}
\label{eq:alpha_12}
\alpha_1=\sqrt{-8q+7}, \qquad \alpha_2=\sqrt{8q-3}
\end{equation}
For $q\in\mathbb{Z}$ and $g(\vartheta)=\cos w$ as in \eqref{eq:ansatz_g_i}, these are expressed as
\begin{align}
&f_{11}(r)=\mathit{c_1}r^{1-\frac{\beta_1}{6}} \mathrm{hypergeom}\! \left(\left[\frac{1}{4}-\frac{\beta_1}{12},\frac{3}{4}-\frac{\beta_1}{12}\right],
\left[1-\frac{\beta_1}{6},1-\frac{\beta_1}{12}-\frac{\beta_2}{12},1-\frac{\beta_1}{12}+\frac{\beta_2}{12}\right],\frac{\left(3 \gamma +4\right)\cdot r^{2}}{12 k}\right)\nonumber,
\end{align}
\begin{align}
&f_{12}(r)=\mathit{c_2}r^{1+\frac{\beta_1}{6}} \mathrm{hypergeom}\! \left(\left[\frac{1}{4}+\frac{\beta_1}{12},\frac{3}{4}+\frac{\beta_1}{12}\right],\left[1+\frac{\beta_1}{6},1+\frac{\beta_1}{12}+\frac{\beta_2}{12},1+\frac{\beta_1}{12}-\frac{\beta_2}{12}\right],\frac{\left(3 \gamma +4\right) r^{2}}{12 k}\right)\nonumber,
\end{align}
\begin{align}
&f_{13}(r)=\mathit{c_3}r^{1-\frac{\beta_2}{6}} \mathrm{hypergeom}\! \left(\left[\frac{1}{4}-\frac{\beta_2}{12},\frac{3}{4}-\frac{\beta_2}{12}\right],\left[1-\frac{\beta_2}{6},1-\frac{\beta_1}{12}-\frac{\beta_2}{12},1+\frac{\beta_1}{12}-\frac{\beta_2}{12}\right],\frac{\left(3 \gamma +4\right) r^{2}}{12 k}\right)\nonumber,
\end{align}
\begin{align}
&f_{14}(r)=\mathit{c_4} r^{1+\frac{\beta_2}{6}} \mathrm{hypergeom}\! \left(\left[\frac{1}{4}+\frac{\beta_2}{12},\frac{3}{4}+\frac{\beta_2}{12}\right],\left[1+\frac{\beta_2}{6},1+\frac{\beta_1}{12}+\frac{\beta_2}{12},1-\frac{\beta_1}{12}+\frac{\beta_2}{12}\right],\frac{\left(3\gamma +4\right) r^{2}}{12 k}\right),\nonumber
\end{align}
with
\begin{equation}
\label{eq:beta_12}
\beta_1=\sqrt{54-3\sqrt{64q^{2}-48q+153}-48q}, \qquad \beta_2=\sqrt{54+3\sqrt{64q^{2}-48q+153}-48q}.
\end{equation}
Upon substituting these solutions into the boundary conditions \eqref{eq:bound_1}-\eqref{eq:bound_4}, we obtain a system of four equations with four unknowns, denoted $\mathit{c_j}$, with $j=1,\cdots.4$. The parameter $k$ in \eqref{eq:k_gamma_elastomeric} is considered a control parameter. When the determinant of the matrix associated with the system is non-zero, the unique solution to the system is the trivial solution, i.e., $f_1\equiv 0$, and so the second variation of $\mathcal{F}$ vanishes only in correspondence of the null-perturbation. If, however, there exists a value of $k$ for which the determinant vanishes, then the system admits infinitely many solutions for $f_1$, defined up to multiplicative constants. In this case, we select the real part of these solutions.



\begin{thebibliography}{23}

\bibitem{bladon:deformation} Bladon P, Terentjev EM, Warner M. 1994. Deformation-induced orientational transitions in liquid crystals elastomer, Journal of Physics II France 4(1), 75–91 (doi:10.1051/jp2:1994100). 

\bibitem{blanch:turbulent} Blanch-Mercader C, Yashunsky V, Garcia S, Duclos G, Giomi L, Silberzan P. 2018. Turbulent dynamics of epithelial cell cultures, Physical Review Letters 120, 208101 (doi:10.1103/PhysRevLett.120.208101).

\bibitem{bouck2024:reduced} Bouck L, Ricardo HN, Shuo Y. 2024. Reduced membrane model for liquid crystal polymer networks: Asymptotics and computation, Journal of the Mechanics and Physics of Solids 187, 105607(doi:10.1016/j.jmps.2024.105607).

\bibitem{brown1973:structure} Brown GH. 1973. Structure, properties, and some applications of liquid crystals, 
Journal of the Optical Society of America, 63, 1505-1514 (doi: 10.1364/JOSA.63.001505). 

\bibitem{cesana2015:effective} Cesana P, Plucinsky P, Bhattacharya K. 2015. Effective behavior of nematic elastomer membranes, Archive for Rational Mechanics and Analysis 218, 863–905 (doi:10.1007/s00205-015-0871-0).

\bibitem{Cirak:2014:CLBW} Cirak F, Long Q, Bhattacharya K, Warner M. 2014. Computational analysis of liquid crystalline elastomer membranes: Changing Gaussian curvature without stretch energy, International Journal of Solids and Structures 51(1), 144-153 (doi: 10.1016/j.ijsolstr.2013.09.019).

\bibitem{conti2018:adaptive} Conti S, Dolzmann G. 2018. An adaptive relaxation algorithm for multiscale problems and application to nematic elastomers, Journal of the Mechanics and Physics of Solids 113, 126–143 (doi:10.1016/j.jmps.2018.02.001).

\bibitem{copenhagen:topological} Copenhagen K, Alert R, Wingreen N, Shaevitz J. 2021. Topological defects promote layer formation in myxococcus xanthus colonies, Nature Physics 17, 211–215 (doi:10.1038/s41567-020-01056-4211).

\bibitem{degennes:physics} De Gennes PG, Prost J. 1993. The Physics of Liquid Crystals
(The International Series of Monographs on Physics), vol 83 2nd edn, Clarendon, Oxford.

\bibitem{duffy2020:defective} Duffy D, Biggins JS. 2020. Defective nematogenesis: Gauss curvature in programmable shape-responsive sheets with topological defects, Soft Matter 16, 10935 (doi: 10.1039/D0SM01192D).

\bibitem{ericksen:liquid} Ericksen JL. 1991. Liquid crystals with variable degree of orientations, Archive for Rational Mechanics and Analysis 113, 97–120 (doi:10.1007/BF00380413).

\bibitem{fernandez2013:liquid} Fernandez G. 2023. Exotic actuators, Nature Materials 12, 12  (doi:10.1038/nmat3526).

\bibitem{Fried:2001:FT} Fried E, Todres RE. 2001. Prediction of disclinations in nematic elastomers, Proceedings of the National Academy of Sciences (PNAS) 98(26), 14773-14777 (doi: 10.1073 pnas.261395098).
	
\bibitem{Fried:2002:FT} Fried E, Todres RE. 2002. Disclinated states in nematic elastomers, Journal of the Mechanics and Physics of Solids 50(12), 2691-2716 (doi: 10.1016/S0022-5096(02)00013-3).

\bibitem{fried2002:normal} Fried E, Todres, RE. 2002. Normal-stress differences and the detection of disclinations in nematic elastomers, Journal of Polymer Science part B-Polymer Physics 40, 2098-2106 (doi:10.1002/polb.10257).

\bibitem{gruzdenko2022:liquid} Gruzdenko A, Dierking I. 2022. Liquid crystal-based actuators, Frontiers in Soft Matter 2, 1052037 (doi:10.3389/frsfm.2022.1052037).

\bibitem{kawaguchi:topological} Kawaguchi K, Kageyama R, Sano M. 2017. Topological defects control collective dynamics in neural progenitor cell cultures, Nature 545, 327-331 (doi:10.1038/nature22321).

\bibitem{keber:topology} Keber FC, Loiseau E, Sanchez T, DeCamp SJ, Giomi L, Bowick MJ, Marchetti MC, Dogic Z, Bausch AR. 2024. Topology and dynamics of active nematic vesicles, Science 345, 1135-1139 (doi:10.1126/science.1254784).

\bibitem{kralj:curvature} Kralj S, Rosso R, Virga E. 2011. Curvature control of valence on nematic shells, Soft Matter 7, 670-683 (doi:10.1039/C0SM00378F).

\bibitem{kumar:tunable} Kumar N, Zhang R, de Pablo JJ, Gardel ML. 2018. Tunable structure and dynamics of active liquid crystals, Science Advances 4, eaat7779 (doi:10.1126/sciadv.aat7779).

\bibitem{Long:2021:etal} Long C, Tang X, Selinger RBL, Selinger JV. 2021. Geometry and mechanics of disclination lines in 3D nematic liquid crystals, Soft Matter 17, 2265-2278 (doi:  10.1039/d0sm01899f).

\bibitem{Maroudas-Sacks:2021:etal} Maroudas-Sacks Y, Garion L, Shani-Zerbib L, Livshits A, Braun E, Keren K. 2021. Topological defects in the nematic order of actin fibres as organization centres of Hydra morphogenesis, Nature Physics 17, 251–259 (doi: 10.1038/s41567-020-01083-1).

\bibitem{McConney:2013:etal} McConney ME, Martinez A, Tondiglia VP, Lee KM, Langley D, Smalyukh II, White TJ. 2013. Topography from topology: Photoinduced surface features generated in liquid crystal polymer networks, Advanced Materials 25, 5880-5885 (doi: 10.1002/adma.201301891).

\bibitem{McCracken:2020:etal} McCracken JM, Donovan BR, White TJ. 2020. Materials as machines, Advanced Materials 32, 1906564 (doi: 10.1002/adma.201906564). 

\bibitem{Mihai:2022} Mihai LA. 2022. Stochastic Elasticity: A Nondeterministic Approach to the Nonlinear Field Theory, Springer Cham, Switzerland (doi: 10.1007/978-3-031-06692-4).

\bibitem{Mihai:2020b:MG} Mihai LA, Goriely A. 2020. A plate theory for nematic liquid crystalline solids, Journal of the Mechanics and Physics of Solids 144, 104101 (doi: 10.1016/j.jmps.2020.104101).

\bibitem{modes2010:disclination} Modes CD, Bhattacharya K, Warner M. 2010. Disclination-mediated thermo-optical response in nematic glass sheets, Physical Review E 81, 060701 (doi:10.1103/PhysRevE.81.060701).

\bibitem{modes2011:gaussian} Modes CD, Bhattacharya K., Warner M. 2011. Gaussian curvature from flat elastica sheets, Proceedings of the Royal Society A 467, 1121 (doi:10.1098/rspa.2010.0352

\bibitem{narayan:long} Narayan V, Ramaswamy S, Menon N. 2007. Long-lived giant number fluctuations in a swarming granular nematic, Science 317, 105-108 (doi:10.1126/science.1140414).

\bibitem{ozenda2020:blend} Ozenda O, Sonnet A, Virga EG. 2020. A blend of stretching and bending in nematic polymer networks, Soft Matter 16, 8877 (doi:10.1039/D0SM00642D).

\bibitem{paparini:shape} Paparini S, Virga EG. 2021. Shape bistability in 2D chromonic droplets, Journal of Physics: Condensed Matter 33(49), 495101 (doi:10.1088/1361-648X/ac2645).

\bibitem{paparini:spiralling} Paparini S, Virga EG. 2023. Spiralling defect cores in chromonic hedgehogs, Liquid Crystals 50(7–10), 1498–1516 (doi:10.1080/02678292.2023.2190626).

\bibitem{paparini:stability} Paparini S, Virga EG. 2022. Stability against the odds: the case of chromonic liquid crystals, Journal of Nonlinear Science 32, 74 (doi:10.1007/s00332-022-09833-6).

\bibitem{saw:topological} Saw TB, Doostmohammadi A, Nier V, Kocgozlu L, Thampi S, Toyama Y, Marcq P, Lim C, Yeomans J, Ladoux B. 2024. Topological defects in epithelia govern cell death and extrusion, Nature 544, 212–216 (doi:10.1038/nature21718).

\bibitem{singh2023:bending} Singh H, Virga EG. 2023. Bending and stretching in a narrow ribbon of nematic polymer networks, Journal of Elasticity 154, 531–553 (doi:10.1007/s10659-022-09978-1).

\bibitem{Selinger:2024} Selinger JV. 2024. Introduction to Topological Defects and Solitons: In Liquid Crystals, Magnets, and Related Materials, Springer Cham, Switzerland (doi: 10.1007/978-3-031-70200-6). 

\bibitem{sonnet2022:photoresponsive} Sonnet AM, Virga EG. 2024. Model for a photoresponsive nematic elastomer ribbon, Journal of Elasticity 155, 327–354 (doi:10.1007/s10659-022-09959-4).

\bibitem{Tang:2017:TS} Tang X, Selinger JV. 2017. Orientation of topological defects in 2D nematic liquid crystals, Soft Matter 13, 5481-5490 (doi: 10.1039/c7sm01195d).

\bibitem{Terentjev:2025} Terentjev EM. 2025. Liquid crystal elastomers: 30 years after, Macromolecules 58, 2792-2806 (doi: 10.1021/acs.macromol.4c01997).

\bibitem{Treloar:1944} Treloar LRG. 1944. Stress-strain data for vulcanized rubber under various types of deformation, Transactions of the Faraday Society 40, 59-70 (doi: 10.1039/TF9444000059).

\bibitem{Warner:2020} Warner M. 2020. Topographic mechanics and applications of liquid crystalline solids, Annual Review of Condensed Matter Physics 11, 125-145 (doi: 10.1146/annurev-conmatphys-031119-050738).

\bibitem{warner1988:theory} Warner M, Gelling K P, Vilgis TA. 1988. Theory of nematic networks, The Journal of Chemical Physics 88(6), 4008–4013 (doi:10.1063/1.453852).

\bibitem{Warner:2007:WT} Warner M, Terentjev EM. 2007. Liquid Crystal Elastomers, paper back, Oxford University Press, Oxford, UK.

\bibitem{warner1991:elasticity} Warner M, Wang X. 1991. Elasticity and phase behavior of nematic elastomers, Macromolecules 24, 4932-4941 (doi: 10.1021/ma0 0 017a033).

\bibitem{White:2018} White TJ. 2018. Photomechanical effects in liquid crystalline polymer networks and elastomers, Journal of Polymer Science, Part B: Polymer Physics 56, 695-705 (doi: 10.1002/polb.24576).

\bibitem{White:2015:WB} White TJ, Broer DJ. 2015. Programmable and adaptive mechanics with liquid crystal polymer networks and elastomers, Nature Materials 14, 1087-1098 (doi: 10.1038/nmat4433). 

\bibitem{Xiao:etal:2024} Xiao T, Wu J, Zhang Y. 2024. Recent advances in the design, fabrication, actuation mechanisms and applications of liquid crystal elastomers, Soft Science 3, 11 (doi: 10.20517/ss.2023.03).

\end{thebibliography}
\end{document}